\documentstyle[emulateapj,psfig]{article}

\slugcomment{To appear in {\it the Astrophysical Journal}}
\lefthead{KOBAYASHI \& ARIMOTO}
\righthead{Line Strength Gradients in Ellipticals}

\def\gtsim {>\kern-1.2em\lower1.1ex\hbox{$\sim$}~}   
\def\ltsim {<\kern-1.2em\lower1.1ex\hbox{$\sim$}~}   

\begin{document}

\title{Gradients of Absorption Line Strengths in Elliptical Galaxies}

\author{Chiaki KOBAYASHI}
\affil{Department of Astronomy, School of Science, University of Tokyo \\
	7-3-1, Hongo, Bunkyo-ku, Tokyo 113-0033, Japan; \\
      e-mail: chiaki@astron.s.u-tokyo.ac.jp}
\author{Nobuo ARIMOTO}
\affil{Institute of Astronomy, School of Science, University of Tokyo \\
        2-21-1, Osawa, Mitaka, Tokyo 181-8588, Japan; \\
      e-mail: arimoto@mtk.ioa.s.u-tokyo.ac.jp}

\begin{abstract}

We have re-studied line-strength gradients of 80 elliptical galaxies. 
Typical metallicity gradients of elliptical galaxies are
$\Delta {\rm [Fe/H]}/ \Delta \log r \simeq -0.3$
which is flatter than 
the gradients predicted by monolithic collapse simulations.
The metallicity gradients do not correlate with 
any physical properties of galaxies, 
including central and mean metallicities, 
central velocity dispersions $\sigma_0$, absolute B-magnitudes $M_B$, 
absolute effective radii $R_{\rm e}$, and dynamical masses of galaxies.
By using the metallicity gradients, 
we have calculated mean stellar metallicities
for individual ellipticals. 
Typical mean stellar metallicities are
$\langle {\rm [Fe/H]} \rangle \simeq -0.3$, and range from 
$\langle {\rm [Fe/H]} \rangle \simeq -0.8$ to $+0.3$,
which is contrary to what Gonzalez \& Gorgas (1996) claimed; 
the mean metallicities of ellipticals are not universal.
The mean metallicities correlate well with $\sigma_0$ and dynamical masses, 
though relations for M$_B$ and R$_{\rm e}$ include significant scatters.
We find fundamental planes defined by surface brightnesses SB$_{\rm e}$, 
$\langle {\rm [Fe/H]} \rangle$, and $R_{\rm e}$ (or $M_B$), and 
the scatters of which are much smaller than those of the 
$\langle {\rm [Fe/H]} \rangle$ - $R_{\rm e}$ (or $M_B$) relations.
The $\langle [{\rm Fe/H]} \rangle$ - $\log \sigma_0$ 
relation is nearly in parallel to the [Fe/H]$_0$ - $\log\sigma_0$ 
relation but systematically lower by 0.3 dex; thus 
the mean metallicities are about a half of the central values. 
The metallicity-mass relation, or equivalently, the color-magnitude relation
of ellipticals holds not only for the central part but also 
for the whole part of galaxies.
Assuming that Mg$_2$ and Fe$_1$ give [Mg/H] and [Fe/H], respectively, 
we find $\langle {\rm [Mg/Fe]} \rangle \simeq +0.2$ in most of 
elliptical galaxies. 
$\langle {\rm [Mg/Fe]} \rangle$ shows no correlation with galaxy mass tracers
such as $\sigma_0$, 
in contrast to what was claimed for the central [Mg/Fe]. 
This can be most naturally explained if the star formation had stopped 
in elliptical galaxies before the bulk of Type Ia supernovae began to explode.
Elliptical galaxies can have significantly different metallicity gradients and 
$\langle \rm{[Fe/H]} \rangle$ even if they have the same galaxy mass.
This may result from galaxy mergers, 
but no evidence is found from presently available data 
to support the same origin for metallicity gradients, 
the scatters around metallicity-mass relation,
and dynamical disturbances.
This may suggest that the scatters have their origin 
at the formation epoch of galaxies.

\end{abstract}

\keywords{galaxies: abundances --- galaxies: elliptical --- galaxies:
evolution --- galaxies: formation}

\section {INTRODUCTION}

How elliptical galaxies formed is one of key questions of modern astronomy. 
Two competing scenarios have so far been proposed: 
Elliptical galaxies should form monolithically by gravitational collapse
of gas cloud with considerable energy dissipation 
(e.g., Larson 1974b; Arimoto \& Yoshii 1987), or alternatively 
ellipticals should form via mergers of relatively small galaxies 
(e.g., Toomre \& Toomre 1972; Kauffmann \& White 1993; Cole et al. 1994). 
Elliptical galaxies show apparently little evidence for on-going star
formation, the bulk of their stars are old (e.g., Kodama \& Arimoto 1997; 
Stanford, Eisenhardt, \& Dickinson 1997; Kodama et al. 1998a), 
and yet some of ellipticals show strong signs of recent dynamical disturbances 
(Schweizer et al. 1990; Schweizer \& Seitzer 1992).
Observational evidences are confusing and controversial.

The merger hypothesis assumes that gaseous disk-like galaxies formed first by 
assembling subgalactic clumps, then two grown-up disk galaxies of similar
mass collided into a single giant elliptical galaxy 
(e.g., Kauffmann \& White 1993; Kauffman \& Charlot 1998). 
Alternatively, an elliptical galaxy could form if many 
small galaxies accreted onto a massive disk galaxy 
(e.g., Cole et al. 1994; Baugh et al. 1998). 
Observationally, the dynamical disturbances such as
shells/ripples, transient dust lanes, 
and multiple and/or counter-rotating cores 
(e.g., Kormendy \& Djorgovski 1989) 
all seem to support the scenario that elliptical galaxies 
formed via hierarchical clustering of smaller galaxies 
(Schweizer \& Seitzer 1992), 
which is a picture predicted by cold dark matter (CDM) cosmology. 
The merger hypothesis may easily explain 
morphology-density relation of galaxies in clusters 
(Dressler 1980; Dressler et al. 1997), 
a significant number of interacting galaxies at high redshifts 
(Driver, Windhorst, \& Griffiths 1995), 
Butcher-Oemler effects (Butcher \& Oemler 1978; 1984), 
and E+A galaxies (Dressler \& Gunn 1983; 1992).
However, collisions and mergers of galaxies should be 
less frequent at the cluster center 
because of large velocity dispersions of galaxies there (Ostriker 1980).
A recent HST observation reveals conspicuous spiral arms dominated by
A-type stars in the E+A galaxies (Franx et al. 1996), 
which suggests that the Butcher-Oemler galaxies are not ellipticals, 
but spiral galaxies falling onto a cluster potential for the first time 
and are witnessed at an instance of being transformed into S0 galaxies.
Elliptical galaxies are surrounded by a huge number of globular clusters (GCs).
The number of globular clusters per galaxy luminosity (specific frequency)
is almost twice of that in spiral galaxies (Harris 1991). 
If ellipticals formed via mergers of spiral galaxies,
at least similar amount of GCs should be born during the merger events
(van den Bergh 1982; 1990).
True, young GCs are discovered in merging galaxies 
(Whitmore \& Schweizer 1995), but these GCs are far less numerous 
than those required for explaining high specific frequencies of ellipticals.
On the other hand, 
GCs in elliptical galaxies have bimodal [Fe/H] distributions 
(Forbes, Brodie, \& Grillmair 1997), 
which has been considered to support the merging hypothesis.

The dissipative collapse hypothesis assumes that a bulk of stars in
ellipticals formed during an initial burst of star formation, 
which was induced by the collisions of fragmented clouds in proto-galaxies, 
and terminated by a supernovae-driven galactic wind that expelled 
the left-over interstellar gas from galaxies 
(e.g., Larson 1974b; Arimoto \& Yoshii 1987). 
The galactic wind is supposed to play an essential role 
in enriching heavy elements in hot intracluster gas (Ciotti et al. 1991). 
Elliptical galaxies after the wind should evolve passively 
(Kodama et al. 1998a). 
The shells and ripples would appear when ellipticals captured 
nearby dwarf galaxies. These dynamical disturbances would be detectable for
a couple of Gyrs, but the capture itself would not introduce any 
significant change in the 
stellar constituents of galaxies, since the mass involved in the secondary 
formation of stars is at most 10\% as the study of H$\beta$ absorption and 
broad band colors of elliptical galaxies shows (Kodama \& Arimoto 1998). 
The galactic wind predicts tight correlations among global properties of
galaxies including a color-magnitude relation (Bower, Lucey, \& Ellis 1992) 
and a fundamental plane (Djorgovski \& Davies 1987; Dressler et al. 1987). 
Recent observations of clusters at high redshifts confirm that 
these relationships are holding even at $z \sim 1$ 
(Dickinson 1996; Schade, Barrientos \& Lopez-Cruz 1997; Kelson et al. 1997; 
Stanford et al. 1998). A progressive change of the color-magnitude
relation as a function of look-back time clearly indicates 
that a bulk of stars in ellipticals formed at the redshift 
$z_{\rm f} \gtsim 2.5-4$ (Kodama et al. 1998a).

In this paper, we study gradients of absorption line strengths of elliptical 
galaxies. In the last two decades, the line strength gradients of ellipticals 
were extensively studied by 
Faber (1977; F77), 
Efstathiou \& Gorgas (1985; EG), 
Couture \& Hardy (1988; CH), 
Peletier (1989, PEL), 
Gorgas, Efstathiou, \& Arag\'on-Salamanca (1990; GEA), 
Boroson \& Thompson (1991; BT), 
Davidge (1991a; D91a), 
Davidge (1991b; D91b), 
Bender \& Surma (1992; BS), 
Davidge (1992, D92), 
Carollo, Danziger, \& Buson (1993; CDB), 
Davies, Sadler, \& Peletier (1993; DSP), 
Gonzalez (1993; GON), 
Hes \& Peletier (1993; HP), 
Saglia et al. (1993; SAG), 
Carollo \& Danziger (1994a; CDa), 
Carollo \& Danziger (1994b; CDb), 
Sansom, Peace, \& Dodd (1994; SPD), 
Cardiel, Gorgas, \& Arag\'on-Salamanca (1995; CGA), 
and Fisher, Franx, \& Illingworth (1995b; FFI). 
Major lines so far studied are Mg$_2$, Mg$_{\rm b}$, 
Fe$_1$(5270$\AA$), Fe$_2$(5335$\AA$), H${\beta}$, and H${\gamma}$ lines.
 The first four metallic lines are degenerate in age and metallicity, while
the two hydrogen lines are sensitive to stellar age 
(Burstein et al. 1984; Faber et al. 1985; Gonzalez 1993; Worthey 1994). 
Recently, Vazdekis \& Arimoto (1999) 
have broken up the age-metallicity degeneracy of 
H${\gamma}$ by including neighboring metallic lines in such a way that a 
resulting H${\gamma}$ depends only on age. 
The approach is thus very promising, but no attempt has yet been 
conducted for estimating ages of giant ellipticals. We therefore should
keep it 
in our mind that some of giant elliptical galaxies in the field might be 
significantly younger than cluster ellipticals 
(Franceschini et al. 1998; Kodama, Bower, \& Bell 1998b) 
and that the radial gradients 
of metallic lines may be caused by not only a radial variation of stellar
metallicity but also an equivalent variation of mean stellar age. 
This paper discusses the gradients of the four metallic lines and H${\beta}$, 
but mainly Mg$_2$ and Mg$_{\rm b}$ gradients, because Fe$_1$ and Fe$_2$ are 
easily influenced by a possible gradient of stellar velocity dispersion whose
spatial distribution in a galaxy is still difficult to measure precisely.

Previous studies showed that (Mg$_2$)$_0$ correlate with the velocity 
dispersion $\sigma_0$ at the galaxy center (Davies et al. 1987; 
Burstein et al. 1988; Bender, Burstein, \& Faber 1993). 
A similar relation holds for 
(Mg$_2$)$_0$ and total absolute magnitudes $M_B$ (Faber 1973; Burstein 1979; 
Terlevich et al. 1981; Dressler 1984). If, {\it admittedly a big if}, one can
assume that elliptical galaxies are old, one can convert Mg$_2$ into the
metallicity $Z$ either empirically (Burstein et al. 1984; Faber et al. 1985)
or theoretically with a help of population synthesis models 
(Mould 1978; Peletier 1989; Buzzoni, Gariboldi, \& Mantegazza 1992; 
Barbuy 1994; Worthey 1994; Casuso et al. 1996; 
Bressan, Chiosi, \& Tantalo 1996; Kodama \& Arimoto 1997). 
For example, if Worthey's (1994) calibration is adopted, one finds that 
a {\it typical} metallicity gradient of elliptical galaxies is 
$\Delta \log Z / \Delta \log r \simeq -0.3$. 
The radial gradient of metallic line strength can be naturally explained
by the dissipative collapse picture.
However, the measured gradients are less steep than those predicted by
numerical simulations of collapse model: 
For example, Larson's hydrodynamical 
simulations gave $\Delta \log Z / \Delta \log r \sim -0.35$ (Larson 1974a), 
and $-1.0$ (Larson 1975); Carlberg's N-body simulations gave 
$\Delta \log Z / \Delta \log r \sim -0.5$ (Carlberg 1984).
This discrepancy could be interpreted if mergers make 
an original gradient flattened; indeed numerical 
simulations showed that the gradient in a disk galaxy should be halved after 
three successive mergers of galaxies with similar size (White 1980).
However, both simulations of the dissipative collapse and successive mergers
leave room for the improvements, 
because some essential physical processes such as 
star formation, thermal feedback from supernovae, and metal
enrichment were not taken into account.

Elliptical galaxies with larger values of (Mg$_2$)$_0$ tend to show 
steeper Mg$_2$ gradients (Gorgas et al. 1990; Carollo et al. 1993; 
Gonzalez \& Gorgas 1996). Since a brighter elliptical usually has a larger 
(Mg$_2$)$_0$ (Davies \& Sadler 1987), Gorgas \& Gonzalez (1996) suggested 
that a larger (Mg$_2$)$_0$ is cancelled by a steeper gradient and that 
elliptical galaxies should have very similar mean stellar metallicities,
regardless of their luminosities and masses. If this is true, the 
color-magnitude relation is nothing but a {\it local} relation that 
holds only for the central part of a galaxy and may not reflect the formation
process of ellipticals. One should use global metallicities instead of
local ones
in studying the origin of ellipticals. Therefore, it is particularly important
to estimate accurately mean stellar metallicities of ellipticals and 
study relations with other global features of galaxies.

No tight correlations between the Mg$_2$ gradient and other properties of 
galaxies are yet confirmed. The Mg$_2$ gradient and the central velocity 
dispersion $\sigma_0$ may correlate (Davidge 1992) but with a significant 
scatter (Gorgas et al. 1990), and cD galaxies deviate from that relation  
considerably (Carollo et al. 1993). However, Davies et al. (1993) did not 
find any significant correlation between the Mg$_2$ gradient and $\sigma_0$.
Nevertheless, the local colors and the line-strengths correlate tightly 
(Davies et al. 1993; Carollo \& Danziger 1994a) and both show strong 
correlation to the local escape velocity (Franx \& Illingworth 1990; 
Davies et al. 1993; Carollo \& Danziger 1994a). Elliptical galaxies which
deviate
considerably from the ridge line of Mg$_2$ -- absolute magnitude relation 
tend to have larger values of $\Sigma$, a degree of dynamical 
disturbances (Schweizer et al. 1990), but the Mg$_2$ gradient is not 
correlated to $\Sigma$, nor the isophotal shape $a_4/a$ (Davidge 1992).

If elliptical galaxies formed monolithically from a massive gas cloud,
the Mg$_2$ gradient should correlate with global properties of galaxies. 
Carlberg (1984) showed that more massive galaxies have steeper metallicity
gradients.
Apparent lack of such correlations may suggest that the Mg$_2$ gradient 
was built up via a series of successive mergers of gas-rich galaxies.
The situation surrounding Mg$_2$ gradient studies is complicated and 
confusing. This we believe is due to a lack of suitable sample of the
Mg$_2$ gradient. A total exposure time required to measure the Mg$_2$
gradient is much longer than that for broad band color gradients.  
Therefore, a single group of authors can observe only a limited number of 
galaxies. What is worse, a quality of data is not uniform. 
We have studied 187 line strength gradients of 133 early-type galaxies.
After carefully examining the quality of data, we have chosen data for
80 elliptical galaxies. Section 2 describes how the data are 
selected and analyzed.
Section 3 gives a definition of the mean stellar metallicities of
ellipticals. Section 4 gives our main results, and our 
discussions and conclusions are given in section 5 and section 6,
respectively.

\section {LINE STRENGTH GRADIENTS}

\subsection{Line Strength Gradients}

 We have studied 187 gradients of Mg$_2$, Mg$_{\rm b}$, Fe$_1$ (5270$\AA$), 
Fe$_2$ (5335$\AA$), and H$\beta$ for 133 early-type galaxies. Data are 
taken from 20 data sources; F77, EG, CH, PEL, GEA, BT, D91a, D91b, BS, 
D92, CDB, DSP, GON, HP, SAG, CDa, CDb, SPD, CGA, and FFI.
To demonstrate a global view of line-strength gradients, we summarize 
Mg$_2$ and Mg$_{\rm b}$ gradients of these galaxies in Appendix A.
Unfortunately, accurate and precise line-strength gradient data are 
difficult to obtain, requiring long integrations on bright objects with
large telescopes to obtain high-enough S/N ratio (e.g., DSP; GON);
thus qualities of
these data are not uniform, and we are obliged to abandon many of them due
to poor qualities as described below (we are indebted to the anonymous
referee 
for reminding us different qualities of the gradient data, and are helped
considerably with the referee's comments in making the following brief
descriptions for the data sources, but we are entirely responsible for any
critical remarks we made): 
{\bf F77} -- poor data quality, no errorbars in figures, but quite large,
 the sky subtraction is poor, and perhaps no
velocity dispersion correction [trustworthiness 
= (index strengths, gradients) = (fair, poor)];
{\bf CH} -- data quality is poor outside nucleus, the gradients are
uncertain, and the sky subtraction is poor [trustworthiness = (poor, poor)];
{\bf PEL} -- the calibration is suspicious, and the errorbars may be 
underestimated [trustworthiness = (good, good)];
{\bf GEA, EG} -- data quality is poor outside nucleus, errors are large, 
the sky subtraction is poor, and the velocity dispersion correction may 
not be done [trustworthiness = (fair, poor)];
{\bf BT} -- the quality of spectra is poor outside nucleus,
the sky subtraction is poor, the errorbars are not given, but
the velocity dispersion correction is reasonable 
[trustworthiness = (fair, good)];
{\bf D91a, D91b, D92} -- the calibrations  
are suspicious, data quality is reasonable for two of three
galaxies, sky subtraction is poor, but it is not clear if velocity dispersion
correction is done or not [trustworthiness 
= (poor, fair?)]; 
{\bf BS} -- the calibration is good, but the sky subtraction is suspicious, 
velocity dispersion corrections? [trustworthiness = (fair, fair)];
{\bf CDB} -- except for Mg$_2$, the calibration is uncertain, significant 
contamination of emissions in galaxy spectra [trustworthiness = (good, 
excellent)]; 
{\bf DSP} -- the calibration is done carefully, the sky subtraction is 
suspicious but better than most [trustworthiness = (excellent, excellent)]; 
{\bf GON} -- the calibration is excellent except for Mg$_1$ and Mg$_2$ which 
are affected by chromatic focus variations, for H$\beta$ the 
velocity dispersion correction is suspicious [trustworthiness = (excellent,
excellent)]; 
{\bf HP} -- the calibration is suspicious [trustworthiness = (good, good)]; 
{\bf SAG} -- zero-points of indices are off, and the velocity 
dispersion correction is suspicious [trustworthiness = (poor, good)];
{\bf CDa, CDb} -- the calibrations are uncertain for Fe$_1$ and Fe$_2$, and
the sky subtraction is suspicious [trustworthiness = (good, excellent)];
{\bf SPD} -- uncalibrated, the sky subtraction is poor, and
the velocity dispersion correction is probably wrong
[trustworthiness = (poor, poor)];
{\bf CGA} -- the calibration is off by 0.014 mag, and the velocity dispersion
correction is not done [trustworthiness = (poor, poor)];
{\bf FFI} -- the calibration is uncertain, the sky subtraction and velocity 
dispersion correction are good, but the emission contamination is observed 
with [OIII] [trustworthiness = (good, excellent)].

Among these 20 data sources, we have adopted 5 references (BS, CDB, DSP, 
CDa, and CDb) that provide excellent Mg$_2$ gradient data for 46 ellipticals
and 2 references (GON and FFI) that give high quality Mg$_{\rm b}$ gradient
data for additional 34 elliptical galaxies. These authors almost
universally used CCDs as detectors instead of photon counters, due to
unreliability of the wavelength and flux calibrations of photon counters
(e.g., Peletier 1989; Trager et al. 1998).

We apply a linear regression analysis for the line-strength gradients by 
taking account of observational errors:
   \begin{equation}
   \mbox{Index}(r) = 
   (\mbox{Index})_{\rm e}+\frac{\Delta\mbox{Index}}
   {\Delta \log r}~\log\frac{r}{r_{\rm e}}
   \equiv A-B\log\frac{r}{r_{\rm e}},
   \end{equation}
where $r$ is a projected radius and a suffix {\it e} indicates 
values defined at an effective radius $r_{\rm e}$.
When the line-strength gradients were measured at different position angles,
we use the effective radius $r_{\rm e}^*$ corrected for an eccentricity,  
as was done by DSP. 
Effective radii and eccentricities are taken from 
the original references, if available, and are supplementally taken from 
Davies et al. (1983), Djorgovski \& Davis (1987), and Davies et al. (1987). 

 In general, linear regression lines give good approximations, 
but observed index values are systematically smaller 
at the inner most central parts and the outer regions beyond $r_{\rm e}$.
At galaxy centers with ${\log r/r_{\rm e}} \ltsim -1.5$, 
the gradients are smeared out due to poor seeing conditions,
and in the outer regions errors arising from the sky subtraction give poor
fits. Therefore, we exclude these regions from our fitting. 
BS, DSP, CDB, CDa, and CDb all used the Lick indices (Burstein et al. 1984;
Faber et al. 1985), while FFI and GON used Gonzalez's (1993) indices.
However, differences are too small to affect the present analysis.
CDB, DSP, CDa, CDb, and FFI converted spectral resolution of their
measurements to Lick scales, but differences are negligible for Mg$_2$ 
and Mg$_{\rm b}$ which are our main concern.

Table 1 gives resulting fitting parameters of the line-strength gradients for 
80 elliptical galaxies. Columns (1)-(7) gives a name of galaxy, reference, 
an index name, an intercept at $r_{\rm e}$ and error, a gradient and error,
respectively. Mg$_{\rm b}$ is most suitable for a study of line-strength
gradient, but the number of galaxies is not large enough to derive the
statistics. The number of galaxies with the Mg$_2$ gradient data is largest 
in our sample. Fe$_1$ shows similar gradients to Mg$_2$,  
but Fe$_2$ shows a larger scatter, since it is very vulnerable to spectral 
resolutions and gradient of velocity dispersions. Therefore, we use Fe$_1$ 
instead of $\langle{\mbox{Fe}}\rangle=(\mbox{Fe}_1+\mbox{Fe}_2)/2$,
which was used in previous studies.
H$\beta$ is an age-sensitive index (Worthey 1994) and 
shows flat or positive gradients (DSP, GON, and FFI).
A negative effect of H$\beta$ emission may explain this (DSP),
but this may be due to a decreasing stellar age 
towards a center of galaxy instead (GON, FFI). If the H$\beta$ gradients
reflect an age gradient within a galaxy, our old-age assumption for
ellipticals is not correct anymore. The metallic line-strength gradients 
(Mg$_2$, Mg$_{\rm b}$, Fe$_1$ and Fe$_2$) are all degenerate in age and  
metallicity, and a simple use of stellar population synthesis model in
converting the line-strength gradients to the metallicity gradients would
not be justified. We have solved the age-metallicity degeneracy 
by analyzing the absorption line gradients together with the H$\beta$
gradients (section 2).

\subsection {Metallicity Gradients}

 Since it is not certain whether all elliptical galaxies are 
equally old, we convert the line-strength gradients to the metallicity 
gradients in two alternative ways: (A) we assume that all ellipticals are 
17 Gyr old and convert the indices to the metallicity, or 
(B) we assume that ellipticals are of different ages 
and the line-strength gradients reflect both the metallicity and age 
gradients. We have solved the age-metallicity degeneracy of the
gradients by studying the Mg$_2$ and H$\beta$ gradients simultaneously,
but the number of galaxies to which H$\beta$ data are available is small, 
and H$\beta$ may be affected by the emission contamination.
Therefore, we first calculate mean stellar metallicities of individual
ellipticals under the assumption (A) and derive correlations 
between the mean stellar metallicity and other physical properties of 
galaxies. We then examine whether scatters of these correlations 
become smaller if we take into account a possible age effect
under the assumption (B).

\subsubsection {Index-Metallicity Relations}

Under the assumption (A), the index-metallicity relations 
can be derived from spectral synthesis models. 
Comparing the relations given by seven different population synthesis
models (Mould 1978; Peletier 1989; Buzzoni et al. 1992; 
Barbuy 1994; Worthey 1994; Casuso et al. 1996; Bressan et al. 1996),
we find that, except Mould (1978) and Barbuy (1994), these 
models give very similar relations unless the metallicity is too
low ([Fe/H] $>-1$). These relations are approximately given by 
the following form: 
\begin{equation}
\mbox{[Fe/H]}=p~\mbox{(Index)}-q.
\end{equation}
Hereafter, we shall use the index-metallicity relations 
derived from Worthey's (1994) SSP models for 17 Gyr old galaxies:
\begin{eqnarray*}
\mbox{[Fe/H]}&=&5.85~\mbox{Mg}_2-1.65,  \\
\mbox{[Fe/H]}&=&0.45~\mbox{Mg}_{\rm b}-1.87,\\
\mbox{[Fe/H]}&=&0.67~\mbox{Fe}_1-2.16, \\
\mbox{[Fe/H]}&=&0.59~\mbox{Fe}_2-1.77.
\end{eqnarray*}
Quite often Mg$_2$ gives the [Fe/H] value $\sim 0.2$ dex higher than that
derived from Fe$_1$. It may be due to our use of 
Worthey's model which assumes $[\alpha/ {\rm Fe}]=0$ for $\alpha$ elements
such as oxygen, magnesium, silicon etc. We will discuss this in section 4.5.
If we use Worthey's SSP models of 12 Gyr, we obtain systematically higher
metallicities, but none of our main conclusions need to be modified.

 If the metallic line-strength gradients are indeed reflecting the metallicity
gradients, the observational data suggest that the gradients can be
approximated
in the following form:
\begin{equation}
\mbox{[Fe/H]}=\log \frac {Z_{\rm e}}{Z_{\odot}} - c \log \frac {r}{r_{\rm e}},
\end{equation}
where $Z_{\odot}=0.02$ is the solar metallicity and 
$Z_{\rm e}$ gives the stellar
metallicity derived at $r=r_{\rm e}$. A coefficient $c$ gives the slope of
metallicity
gradient. On the other hand, from equations (1) and (2) we get the following
equation:
\begin{equation}
\mbox{[Fe/H]}=(Ap-q)-Bp~\log\frac{r}{r_{\rm e}},
\end{equation}
which gives the parameters $Z_{\rm e}$ and $c$ in equation (3) for individual
galaxies.

Figure 1 shows histograms of the line-strength gradients of the four indices, 
$-\Delta {\rm Mg}_2/ \Delta \log r$, 
$-\Delta {\rm Mg}_{\rm b}/ \Delta \log r$, 
$-\Delta {\rm Fe}_1/ \Delta \log r$, and
$-\Delta {\rm Fe}_2/ \Delta \log r$.
Note that there is no sign of bimodality in the frequency distributions
of the Mg$_2$, Mg$_{\rm b}$, Fe$_1$, and Fe$_2$ gradients. 
This implies that, whatever the origin is,
the gradients were formed by a single physical process.
The average values are 
$\Delta {\rm [Fe/H]} / \Delta \log r=-0.30\pm0.12$ (Mg$_2$), 
$-0.34\pm0.16$ (Mg$_{\rm b}$),
$-0.28\pm0.16$ (Fe$_1$), and $-0.25\pm0.13$ (Fe$_2$). 
All indices suggest that the metallicity 
gradients are in the range of 
$-0.6 \le \Delta {\rm [Fe/H]}/ \Delta \log r \le -0.1$. 
A typical metallicity gradient is about 
$\Delta {\rm [Fe/H]} / \Delta \log r \sim -0.3$,
which is slightly steeper
than the original estimate, because we have excluded the central regions
where the gradients are significantly altered by poor seeings. The gradients
thus derived are considerably flatter than a theoretical value $-0.5$
predicted by Carlberg (1984).

Figure 2 shows a frequency distribution of Mg$_2$ at the galaxy center,
(Mg$_2$)$_0$, for 572 elliptical galaxies (dashed line) given by 
Davies et al. (1987), and 80 ellipticals (solid line) studied in this
article.  
We note that there is no systematic difference between the two samples, 
thus we believe that we are dealing with a fairly good sample of 
elliptical galaxies in the local universe. 
Figure 3 gives frequency distributions of the four indices measured at 
the effective radius (i.e., the intercept at $r=r_{\rm e}$). 
As we will see later (section 3), the metallicities evaluated 
at $r=r_{\rm e}$ give good measures for the mean stellar metallicities. 
The average values are 
$\langle {\rm [Fe/H]}\rangle =-0.18\pm0.19$ (Mg$_2$), 
$-0.15\pm0.24$ (Mg$_{\rm b}$),
$-0.37\pm0.17$ (Fe$_1$), and $-0.36\pm0.18$ (Fe$_2$).
Contrary to what
Gonzalez \& Gorgas (1996) claimed, the mean metallicities of ellipticals
are not universal, but range from 
$\langle {\rm [Fe/H]} \rangle \simeq -0.8$ to $+0.3$.
A typical mean stellar metallicity of elliptical galaxies is 
about a half solar, $\langle {\rm [Fe/H]} \rangle \simeq -0.3$,
which is considerably smaller than the stellar 
metallicity [Fe/H]$_0 \simeq +0.2$ measured at the galaxy center. 
From a similar study of the Mg$_2$ gradients but with much smaller 
sample of ellipticals, 
Arimoto et al. (1997) found that the mean metallicity of typical 
elliptical galaxies with $\sigma_0 \ge 250$ km s$^{-1}$ is 
$\langle {\rm [Fe/H]} \rangle \simeq -0.06 \pm 0.13$. 
Our sample includes 19 ellipticals with $\sigma_0 \ge 250$ km s$^{-1}$,
to which we obtain $\langle {\rm [Fe/H]} \rangle \simeq -0.04 \pm 0.11$.
Thus, a claim made by Arimoto et al. (1997) remains unexplained; i.e.,
the iron abundance of the interstellar mediums of luminous elliptical
galaxies, as derived from ASCA X-ray observations of the iron L complex,
is at variance with the abundance expected from the stellar populations
as derived from current population synthesis methods of optical spectrum. 

\subsubsection {Indices-Metallicity-Age Matrices}

 Under the assumption (B), we find that the relations among two 
indices (one is H${\beta}$ and the other is either Mg$_2$ or Mg$_{\rm b}$),
age, and metallicity can be approximated with the following matrix forms:

{\footnotesize
\begin{eqnarray}
\left(\begin{array}{l}
{\rm Index}_1 \\
{\rm Index}_2
\end{array}\right)&=&
\left(\begin{array}{ll}
s_1 & s_2 \\
s_3 & s_4
\end{array}\right)
\left(\begin{array}{l}
t_9 \\
{\rm [Fe/H]}
\end{array}\right)+
\left(\begin{array}{l}
q_1 \\
q_2
\end{array}\right),\\
\left(\begin{array}{l}
t_9 \\
{\rm [Fe/H]}
\end{array}\right)&=&
\left(\begin{array}{ll}
p_1 & p_2 \\
p_3 & p_4
\end{array}\right)
\left(\begin{array}{l}
{\rm Index}_1-q_1 \\
{\rm Index}_2-q_2
\end{array}\right),
\end{eqnarray}
}
where $t_9$ is an age in units of Gyr. If we adopt Worthey's (1994) models,
we obtain :
{\footnotesize
\begin{eqnarray*}
\left(\begin{array}{l}
{\rm H}\beta \\
{\rm Mg}_2
\end{array}\right)&=&
\left(\begin{array}{ll}
-0.0425 & -0.607 \\
0.00414 & 0.182
\end{array}\right)
\left(\begin{array}{l}
t_9 \\
{\rm [Fe/H]}
\end{array}\right)+
\left(\begin{array}{l}
2.12 \\
0.211
\end{array}\right),\\
\left(\begin{array}{l}
t_9 \\
{\rm [Fe/H]}
\end{array}\right)&=&
\left(\begin{array}{ll}
-34.9 & -116. \\
0.792 & 8.12
\end{array}\right)
\left(\begin{array}{l}
{\rm H}\beta-2.12 \\
{\rm Mg}_2-0.211
\end{array}\right),\\
\left(\begin{array}{l}
{\rm H}\beta \\
{\rm Mg}_{\rm b}
\end{array}\right)&=&
\left(\begin{array}{ll}
-0.0425 & -0.607 \\
0.0583 & 2.28
\end{array}\right)
\left(\begin{array}{l}
t_9 \\
{\rm [Fe/H]}
\end{array}\right)+
\left(\begin{array}{l}
2.12 \\
3.18
\end{array}\right),\\
\left(\begin{array}{l}
t_9 \\
{\rm [Fe/H]}
\end{array}\right)&=&
\left(\begin{array}{ll}
-37.1 & -9.89 \\
0.950 & 0.692
\end{array}\right)
\left(\begin{array}{l}
{\rm H}\beta-2.12 \\
{\rm Mg}_{\rm b}-3.18
\end{array}\right),
\end{eqnarray*}
}
which are valid for $8 \ltsim t_9 \ltsim 17$ and 
$-0.5 \ltsim$ [Fe/H] $\ltsim 0.5$ (see figure 4).
With these equations, we can easily derive both metallicity gradient and
age gradient simultaneously from the two line-strengths gradients.
From equations (1) and (6), we obtain the following formal relations
that give the age and metallicity gradients for individual galaxies:
\begin{equation}
t_9=p_1(A_1-q_1)+p_2(A_2-q_2)-(p_1B_1+p_2B_2)~\log\frac{r}{r_{\rm e}},
\end{equation}
\begin{equation}
\mbox{[Fe/H]}=p_3(A_1-q_1)+p_4(A_2-q_2)-(p_3B_1+p_4B_2)~\log\frac{r}{r_{\rm
e}},
\end{equation}
where $A_1, B_1$ are for Index$_1$ and $A_2, B_2$ are for Index$_2$.

 In applying these relations to individual elliptical, however, we find that
resulting ages for several galaxies are not realistic (much larger than 17 Gyr
old). This may probably due to a poor quality of H${\beta}$ gradient data.
We therefore further assume that the age gradients are small and could be 
ignored. We only take into account a possibility that a whole stellar
population
of galaxy is systematically young, that is, 
we estimate a typical age of galaxy with an intercept of H$\beta$ 
at the effective radius. Assuming no age gradients
 (i.e., $s_2 \sim 0, B_1 \sim 0$), we have derived 
the metallicity gradients by using the following equations for 
Mg$_2$ and Mg$_{\rm b}$:

\begin{equation}
t_9 =(A_1-q_1)/s_1,
\end{equation}
\begin{equation}
\mbox{[Fe/H]}=(A_2-q_2-s_3 t_9)/s_4
-B_2/s_4~\log\frac{r}{r_{\rm e}}.
\end{equation}

\section {MEAN STELLAR METALLICITY}

 It is well known that surface 
brightness profiles of elliptical galaxies are not fitted well with 
so-called de Vaucouleurs' law 
(Caon, Capacciolo, \& D'Onofrio 1993; 
Binggeli \& Jerjen 1997; Graham \& Colless 1997).
We, therefore, use Sersic law, which is a generalized de Vaucouleurs' law,
by replacing a power index $1/4$ with $1/n$.
The parameter $n$ correlates with a luminosity of a galaxy;
a bright elliptical has $n=4$ profile while a dwarf elliptical
has exponential $n=1$ profile (Binggeli \& Jerjen 1997).
In this section, we derive the mean stellar metallicity of an
elliptical galaxy by using the observed line-strength gradients and the 
theoretical index - metallicity relations.

We assume that an elliptical galaxy is spherically symmetric and 
the surface brightness profile is given by Sersic law:
\begin{equation}
I(r)=I_{\rm e} \exp \left[-b\left\{\left(\frac{r}{r_{\rm e}}
\right)^\frac{1}{n}-1\right\}\right],
\end{equation}
\begin{equation}
I_{\rm e}=\frac{1}{2n\pi e^b b^{-2n} \Gamma(2n)}\frac{L_*}{r_{\rm e}^2},
\end{equation}
where $r$ is the projected radius and 
$r_{\rm e}$ is de Vaucouleurs' effective radius.
$b$ is a function of $n$ determined from the definition of the 
effective radius and 
$b(n)$ is approximately given by the following equation:
\begin{equation}
b(n)=-0.326463 + 1.99927~n.
\end{equation}
The luminosity within the radius $r$ is given by:
\begin{equation}
L(r)=\int^r_0 2\pi r I(r) dr
=2n\pi e^b b^{-2n} r_{\rm e}^2 I_{\rm e}~
\Gamma\left(2n,0,b\left(\frac{r}{r_{\rm e}}\right)^\frac{1}{n}\right),
\end{equation}
where $\Gamma$ is the general Gamma function.
The total luminosity is given by:
\begin{equation}
L(\infty)=2n\pi e^b b^{-2n} \Gamma(2n) r_{\rm e}^2 I_{\rm e}=L_*.
\end{equation}
If we assume that the projected metallicity distribution 
is given by 
an exponential form, as the observed Mg$_2$ gradients suggest:
\begin{equation}
\bar{Z}(r)=Z_{\rm e}\left(\frac{r}{r_{\rm e}}\right)^{-c},
\end{equation}
the metallicity gradient is then given as:
\begin{displaymath}
\mbox{[Fe/H]}=\log\bar{Z}(r)=\log\frac{Z_{\rm e}}
{Z_\odot}-c\log\frac{r}{r_{\rm e}},
\end{displaymath}
which is the same as equation (3).
The parameters $c$ and $Z_{\rm e}$ are determined from
the parameters $A$ and $B$ in equation (1)
and the index-metallicity relations are given by equation (8), or
equations (4) and (10).

In an analogy to the surface brightness of stars,
the {\it surface brightness profile of metals}
can be defined as:
\begin{equation}
I_Z(r)=\bar{Z}(r) I(r).
\end{equation}
Then the {\it luminosity of metals} within the radius $r$ is given by:
\begin{equation}
L_Z(r)= \int^r_0 2\pi r I_Z(r) dr
=b^{cn} \frac{\Gamma\left(2n-cn,0,b\left(\frac{r}{r_{\rm e}}
\right)^\frac{1}{n}\right)}
{\Gamma\left(2n,0,b\left(\frac{r}{r_{\rm e}}\right)^\frac{1}{n}\right)} 
Z_{\rm e} L_*.
\end{equation}
The {\it total luminosity of metals} is given as follows:
\begin{equation}
L_Z(\infty)=b^{cn} \frac{\Gamma(2n-cn)}{\Gamma(2n)} Z_{\rm e} 
L_* \simeq Z_{\rm e} L_*,
\end{equation}
where we note that the constant $b^{cn} \Gamma(2n-cn)/\Gamma(2n)$ 
is nearly equal to unity as is shown below.

If we assume that a mass-to-light ratio $M/L$ is constant within a galaxy, 
the total mass of metals contained in stars can be given as:
\begin{equation}
M_Z(\infty)=b^{cn} \frac{\Gamma(2n-cn)}{\Gamma(2n)} Z_{\rm e} 
M_* \simeq Z_{\rm e} M_*,
\end{equation}
where $M_*$ is the total mass of stars in a galaxy.
Finally the mean stellar metallicity can be given by:
\begin{equation}
\langle {Z}\rangle \equiv M_Z(\infty)/M^*
=b^{cn} \frac{\Gamma(2n-cn)}{\Gamma(2n)} Z_{\rm e} \simeq Z_{\rm e}.
\end{equation}
Although equation (21) gives an excellent measure for the mean stellar
metallicity of elliptical galaxy, we hereafter use an explicit 
definition to calculate $\langle {\rm [Fe/H]} \rangle$ in such a way 
that we can take into account observed scatters of the metallicity 
gradients:
\begin{equation}
\langle {\rm [Fe/H]} \rangle 
= \log \left[ b^{cn} \frac {\Gamma(2n-cn)}{\Gamma(2n)} 
\frac {Z_{\rm e}}{Z_{\odot}} \right].
\end{equation}

In table 2, we give the metallicity gradient parameters 
for 80 elliptical galaxies derived under the assumption (A)
(hereafter, unless otherwise mentioned, all metallicities are
calculated under the assumption (A)).
Columns (1)-(2) give the name of galaxy and the reference.
Columns (3)-(4), (5)-(6), (7)-(8) and (9)-(10) give the metallicity 
$Z_{\rm e}$ measured at the effective radius and the gradient $c$ 
derived from Mg$_2$, Mg$_{\rm b}$, Fe$_1$, and Fe$_2$ gradients, respectively.
Columns (11)-(15) give the mean metallicities derived from the Mg$_2$,
Mg$_{\rm b}$, Fe$_1$, and Fe$_2$ gradients, respectively.

A typical elliptical galaxy has $n=4$ and $c\sim0.3$ (Figure 1), 
to which the constant $b^{cn}{\Gamma(2n-cn)}/{\Gamma(2n)}$ 
in equation (21) becomes $\simeq 1.1$, thus
the mean metallicity of galaxy is very close to the metallicity measured 
at the effective radius. Arimoto et al. (1997) already showed that this
is the case for smaller sample of galaxies if de Vaucouleurs' surface 
brightness profile is assumed. In this study, we have generalized it to 
Sersic's surface brightness profile and have found that so far as the slope
$c$ is smaller than 0.4, as is the case for most ellipticals,
the mean metallicity is approximately given by $Z_{\rm e}$, which is smaller
than the true mean by at most $\sim$ 0.15 dex. $Z_{\rm e}$ values estimated by
the Mg$_2$-, Mg$_{\rm b}$-, Fe$_1$-, and Fe$_2$-gradient are given in table 2.

Table 3a gives the mean metallicities for 12 ellipticals
derived from the Mg$_2$ gradients under the assumption (B). 
H${\beta}$ corrections are explicitly taken into account.
Columns (1)-(2) give the name of galaxy and the reference.
Columns (3),(5), and (8) give the mean metallicities calculated 
from different equations as are indicated on the top line of the
table; i.e., column (3) gives $\langle {\rm [Fe/H]} \rangle$ without any age
gradient corrections, column (5) gives $\langle {\rm [Fe/H]} \rangle$ 
with constant age corrections, and column (8) gives 
$\langle {\rm [Fe/H]} \rangle$ with age gradient corrections, respectively.
Column (4) gives a typical age $(t_9)_{\rm e}$ estimated from
the intercept of H$\beta$ at the effective radius by using equation (9).
Columns (6)-(7) are parameters of the age gradients, 
the intercept $(t_9)_{\rm e}=p_1(A_1-q_1)+p_2(A_2-q_2)$ 
at the effective radius and 
the gradient $\Delta t_9=-(p_1B_1+p_2B_2)$, both calculated with equation
(7). 
For individual galaxies, $\langle {\rm [Fe/H]} \rangle$ given in columns (5) 
and (8) are nearly identical, thus the age gradient correction is not 
crucial for deriving $\langle {\rm [Fe/H]} \rangle$. For NGC4278, NGC4486, 
NGC5636, NGC4839, and NGC7626 we obtain unexpectedly large $\Delta t_9$. 
This is probably because the H$\beta$ gradients are contaminated by the
emission components. We note that $\langle {\rm [Fe/H]} \rangle$ given in
column (3) are not much different from those given in columns (5) and (8).
Therefore, we conclude that the age corrections for $\langle {\rm [Fe/H]}
\rangle$ 
are not important. Table 3b is the same
as table 3a, but for 31 ellipticals with the Mg$_{\rm b}$ gradients.
The same is true for $\langle {\rm [Fe/H]} \rangle$ given in table 3b.
Except for 16 galaxies to which both equations (10) and (8) give unreasonable
ages and age gradients, the resulting $\langle {\rm [Fe/H]} \rangle$ 
given in columns (3), (5), and (8) are nearly the same.

\section {GLOBAL SCALING RELATIONS}

Gonzalez \& Gorgas (1996) claimed that elliptical galaxies with
larger $({\rm Mg}_2)_0$ tend to have steeper Mg$_2$ gradients.
They therefore suggested that the mass-metallicity relation
of elliptical galaxies holds only for the central metallicity and 
claimed that the mean metallicity of elliptical galaxies should be universal. 
Since the mass-metallicity relation, or equivalently the color-magnitude
relation, 
of elliptical galaxies has been a fundamental relation that any theory
of galaxy formation should account for (e.g., Arimoto \& Yoshii 1987;
Kodama \& Arimoto 1997; Kauffmann \& Charlot 1998), 
it is important to verify if mean
stellar metallicities of ellipticals correlate tightly with 
other global properties, including luminosities,
or if the mean metallicities are just universal for all ellipticals,
regardless of their masses.

Table 4 gives physical properties of 80 elliptical galaxies: 
Columns (1)-(3) gives the name of galaxy, the reference, and the morphology
type
taken from RC3, respectively. Columns (4)-(8) gives the central Mg$_2$ index
(mag), the central velocity dispersion (km s$^{-1}$), both are taken from 
Davies et al. (1987), the absolute
effective radius (kpc), the absolute magnitude 
in B-band (mag), and $\kappa_1=(\log\sigma_0^2+\log R_{\rm e})/\sqrt{2}$
introduced by Bender, Burstein \& Faber (1992) 
as a galaxy mass tracer, respectively.
The effective radius and the absolute magnitude are calculated from 
Burstein et al. (1987) with the distance modulus estimated from the
$D_n-\sigma$ relation of elliptical galaxies by C.Ikuta \& N.Arimoto (private 
communications). 

In the following arguments, we use if necessary, 
a ratio of the rotational velocity to a velocity 
dispersion $(v/\sigma)^*$, 
a parameter $a_4$ which shows the distortion of the 
isophotal contour (Bender et al. 1993), a parameter $\Sigma$ which shows 
the kinematical peculiarity (Schweizer \& Seitzer 1992), and a gas mass 
$M_{\rm gas}$ which is estimated from the {\it ASCA} X-ray observations 
(Matsushita 1997).

In the following sections, we demonstrate the global scaling relations
for elliptical galaxies.
Table 5 gives a result of our attempts for searching the relations. 
Columns (3) gives the number of galaxies used, 
and column (4) gives the correlation coefficient. The scaling laws are given 
in column (5) if the correlation coefficients
are appreciably significant. 

\subsection {Gradients versus Mean Metallicity}

In figures 5a and 5b, we  plot the Mg$_2$ gradient against [Fe/H]$_0$ and
$\langle {\rm [Fe/H]} \rangle$, respectively.
Contrary to what claimed by Gonzalez \& Gorgas (1996), figures 5a
and 5b show no evidence for any correlation. 
A study of Mg$_{\rm b}$ gradients
concludes the same, although not shown here.

\subsection {Gradients versus Mass}

Figure 6a-6d show the Mg$_2$ gradient against galaxy mass tracers; 
the central velocity dispersion $\sigma_0$,
the absolute B-magnitude $M_B$, the absolute effective radius $R_{\rm e}$,
and $\kappa_1=(\log\sigma_0^2+\log R_{\rm e})/\sqrt{2}$.
Contrary to a prediction of numerical simulation (Carlberg 1984),
figures 6 shows no evidence for any correlation.

\subsection {Metallicity-Mass Relation}

The color-magnitude relation of elliptical galaxies is
usually interpreted as that luminous galaxies have higher mean stellar
metallicities (Arimoto \& Yoshii 1987; Kodama \& Arimoto 1997).
Figures 7a and 7c show how $\log \sigma_0$ correlates with 
[Fe/H]$_0$ and $\langle {\rm [Fe/H]} \rangle$, respectively. 
Both are derived from Mg$_2$. 
The $\langle {\rm [Fe/H]} \rangle$ - $\log\sigma_0$ 
relation is well defined and has nearly the same slope to that of the
${\rm [Fe/H]}_0$ - $\log \sigma_0$ relation, but is systematically 0.3 
dex lower; i.e., the mean stellar metallicity is about a half of the central 
metallicity. We note that the age correction by using H$\beta$ changes
little this relation. 
Figure 7b shows an equivalent relation derived from Mg$_{\rm b}$.
Although the slope is similar to that defined by Mg$_2$, the scatters are
quite large.
Figures 7d-7f are the same as figure 7c, but for
the other mass tracers; $M_B$, $R_{\rm e}$, and $\kappa_1$, 
respectively.
We can also find a similar metallicity-mass relation for $\kappa_1$
with a little larger scatter,
and in the case of $M_B$ and $R_{\rm e}$ scatters become quite larger.
This is because the relation with $M_B$ and $R_{\rm e}$ is 
a face-on view of the fundamental plane, which we discuss in section 4.4.
Therefore, we conclude that the metallicity-mass relation of ellipticals hold 
not only for the galaxy center but also for the whole stellar populations
within a galaxy.
When we use the mean metallicity with the correction of age estimated by 
H$\beta$ in Table 3a and 3b, 
the scatters in the metallicity-mass relations do not become smaller.

\subsection {Fundamental Plane of Metallicity-Mass-Surface Brightness}

 A fundamental plane (FP) is a correlation of elliptical galaxies with $2+n$ 
parameters (e.g., Djorgovski \& Davies 1987; Dressler et al. 1987) and is 
a clue to understand formation and evolution of elliptical galaxies.
One possible interpretation of the FP, 
defined by central velocity dispersions $\sigma_0$, 
absolute effective radii R$_{\rm e}$, and 
surface brightnesses within an effective radius SB$_{\rm e}$
attributes it to a correlation of the mass-to-light 
ratio $M/L$ to the total luminosity, or equivalently, to the total galaxy 
mass (e.g., Faber et al. 1987). 
The FP is observed up to $z\sim 0.5$ by 
recent HST observations (Kelson et al. 1997), which can be understood as
an evidence for passive evolution of 
cluster elliptical galaxies since $z<0.5$. 
The FP can be defined for colors or (Mg$_2$)$_0$ 
instead of $\sigma_0$ (de Carvalho \& Djorgovski 1989). 
As we have seen above, $\langle {\rm [Fe/H]} \rangle$
correlates well to $\sigma_0$. Therefore, we study if 
$\langle {\rm [Fe/H]} \rangle$ 
defines new FPs with other global scaling properties. 
Table 6 gives
the FPs of our sample; the first four are already well known FPs and the
last two are the new FPs found in this study.
Figures 7d and 7e 
($\langle {\rm [Fe/H]} \rangle$ against $M_B$ and $R_{\rm e}$, respectively) 
include significant scatters, which is because these figures are 
the face-on view of the FPs 
($\langle {\rm [Fe/H]} \rangle$, SB$_{\rm e}$, and $M_B$ or $R_{\rm e}$).
Even if we include a parameter SB$_{\rm e}$ to the 
$\langle {\rm [Fe/H]} \rangle -\log\sigma_0$ relation (Figure 7c), 
we cannot reduce the scatter, which means figure 7c provides 
the exact edge-on view of the FP.

Figure 8 shows a new fundamental plane defined by 
$\langle {\rm [Fe/H]} \rangle$, $R_{\rm e}$, and SB$_{\rm e}$. 
We interpret it as
the correlation of the {\it metal mass-to-light ratio} 
$M_z/L$ to the total luminosity.
The FP of $\sigma_0 - {\rm SB}_{\rm e} - R_{\rm e}$ is given by:
\begin{equation}
\sigma_0 \propto R_{\rm e}^{0.64}~\langle\Sigma\rangle_{\rm e}^{0.68}.
\end{equation}
The FP of $\langle {\rm [Fe/H]} \rangle - {\rm SB}_{\rm e} - R_{\rm e}$ is:
\begin{equation}
\langle Z \rangle \propto R_{\rm e}^{1.26}~\langle\Sigma\rangle_{\rm e}^{1.38}.
\end{equation}
With the definition of the surface brightness
 $\langle\Sigma\rangle_{\rm e}\equiv2L/\pi R_{\rm e}^2$
 (SB$_{\rm e}\equiv -2.5\log \langle\Sigma\rangle_{\rm e}$), 
the virial theorem $GM/R_{\rm e}\propto\sigma_0^2$, 
and the definition of metallicity
$M_Z/M\equiv\langle Z \rangle$, 
we get the following equations:
\begin{equation}
M/L\propto R_{\rm e}^{-0.42} L^{0.35}.
\end{equation}
\begin{equation}
M_Z/L\propto R_{\rm e}^{-1.91} L^{1.73}.
\end{equation}
where $L$, $M$ and $M_Z$ are the luminosity, the mass, the mass of the metals
of the galaxy, respectively.

\subsection {Magnesium Enrichment?}

It has been quite often claimed that magnesium and perhaps other 
$\alpha$-elements are overabundant in central regions of
elliptical galaxies (Worthey, Faber, \& Gonzalez 1992; Davies et al. 1993; 
Gonzalez 1993; Fisher et al. 1995b). In particular, Fisher et al. (1995b) 
demonstrated that ellipticals with larger $\sigma_0$ tend to have larger
[Mg/Fe]$_0$ (c.f., Trager et al. 1998). 
We have studied if this is also true for the mean abundances of these
elements. 

 Figure 9 compares $\langle {\rm [Fe/H]} \rangle$ given by Mg$_2$ and Fe$_1$, 
respectively. As is clearly seen, $\langle {\rm [Fe/H]} \rangle_{\rm Mg_2}$ 
are higher than $\langle {\rm [Fe/H]} \rangle_{\rm Fe_1}$ by $\sim 0.2$ dex. 
If Mg$_2$ and Fe$_1$ reflect the abundances of magnesium and iron,
respectively,
figures 9 implies that magnesium is enhanced by $\sim 0.2$ dex with
respect to iron in everywhere within a galaxy. We have checked if there
is any systematic difference in the gradients of Mg$_2$ and Fe$_1$,
but fail to find any clear correlation.
We do not find any relation between $\langle {\rm [Mg/Fe]} \rangle$
and dynamical or luminous masses, as shown in table 6.
We also note that in figure 9 there is no 
trend of increasing $\langle {\rm [Mg/Fe]} \rangle$ 
towards higher $\langle {\rm [Fe/H]} \rangle_{{\rm Mg}_2}$; or in other
words, 
$\langle {\rm [Mg/Fe]} \rangle \simeq +0.2$ in most elliptical galaxies
independent
of their mass, in contrast to what was claimed for [Mg/Fe]$_0$.
This can be seen much clearly in figures 10a-10d in which we plot 
$\langle {\rm [Mg/Fe]} \rangle$ against
$\log \sigma_0$, $M_B$, $R_{\rm e}$, and $\kappa_1$, respectively. 
Obviously, these figures do not show any correlation at all.

 However, a word of caution is to be given. The magnesium enhancement 
might not be real. Although Fe$_1$ is less 
fragile to the velocity dispersion gradient than Fe$_2$, the
observational errors of Fe$_1$ are much larger than those of 
Mg$_2$. It is not yet fully understood whether  
Mg$_2$ gives [Mg/H] and Fe$_1$ traces [Fe/H].
Indeed, Mg$_2$ depends strongly on carbon, iron, and several other spices
 (Tripicco \& Bell 1995).
An increase of metallicity increases the opacity and changes the structure 
of stellar atmosphere, as a result of which the indices will be 
strengthened; 
thus Mg$_2$ should also be sensitive to [Fe/H].
Therefore, one should keep in mind that the ratio of Mg$_2$ to Fe$_1$
may not directly give the [Mg/Fe] ratio. 

The point we wish to make here is as follows: 
Previously, it was claimed that more massive ellipticals tend to have 
[Mg/Fe] ratios (e.g., Fisher et al. 1995b). This claim based on the
assumption that Mg$_2$ and Fe$_1$ give [Mg/H] and [Fe/H], respectively,
and used only central values of Mg$_2$ and Fe$_1$ for their analyses.
Under the same assumption, but if $\langle {\rm [Mg/H]} \rangle$ and
$\langle {\rm [Fe/H]} \rangle$ are used instead, there exists no clear
correlation between $\langle {\rm [Mg/Fe]} \rangle$ and the galaxy mass
(figures 10a-10d).
If the assumption we made turns out to be wrong, our conclusion should be 
wrong, but Fisher et al.'s (1995b) argument should also lose its standing
point.

Magnesium is mainly produced by Type II supernovae (SNe II), while 
the bulk of iron is produced by Type Ia supernovae (SNe Ia),
and SNe II and SNe Ia have different lifetimes; 
$10^{6-8}$ yr and a few Gyr, respectively (c.f., Kobayashi et al. 1998).
Thus, the [Mg/Fe] ratio provides crucial constraints on the 
timescale of chemical enrichment in elliptical galaxies. 
Based on the previous findings for the
ratio of Mg$_2$ and Fe$_1$ at the galaxy center, it has been suggested that 
the magnesium enhancement in elliptical galaxies can be explained 
(e.g., Worthey et al. 1992; Matteucci 1994),
if 1) the duration of star formation is shorter than a few Gyr
before SNe Ia start to explode,
or if 2) a slope of initial mass function (IMF) is flatter than 
a slope of the Salpeter IMF to produce more SNe II, 
or if 3) the metal enriched wind induced the selective
mass loss and ejected the iron enriched gas before it is consumed for 
forming next generations of stars (Vader 1986; 1987), 
or if 4) the binary frequency is smaller and 
produces less SNe Ia in ellipticals (Arimoto et al. 1997),  
or if 5) SNe Ia explode less frequently in metal rich environment, 
which is contrary to what is expected theoretically (Kobayashi et al. 1998),
or if 6) the yields of SNe II significantly depend on the metallicity
(Maeder 1992).
In the present study, we find that $\langle {\rm [Mg/Fe]} \rangle \simeq +0.2$
regardless of the galaxy mass. This can be most naturally explained if the
star formation had stopped in elliptical galaxies before the bulk of SNe Ia
began to explode, as was already predicted by a galactic wind model
by Arimoto \& Yoshii (1987).
A recent detailed study of metallicity effects on SNe Ia
by Kobayashi et al. (1998) suggests that 
duration of the star formation in elliptical galaxies 
should be $\sim 1$ Gyr.

\section {Discussions}

\subsection {Dispersion}

 Global metallicities of elliptical galaxies correlate with central
velocity dispersions but with rather large dispersions (figure 7c).
This comes from the fact that the metallicity gradient of elliptical
galaxies is not uniquely determined by the galaxy mass.
Figure 11 shows three sequences of galaxies with various gradients.
All galaxies are normal ellipticals showing no significant peculiarity.
Left panel shows a sequence of 
$\left({\rm Mg}_2\right)_{\rm e}=\mbox{const.}$, 
so that these galaxies have the same intercepts at the 
effective radius and yet show different gradients. This means that 
elliptical galaxies can have the same mean metallicities even if the
metallicities at the galaxy center are considerably different.
Middle panel shows a sequence of 
$\left({\rm Mg}_2\right)_0=\mbox{const.}$, 
thus these galaxies have the same central Mg$_2$, but have  
different gradients. This implies that even if the galaxy mass is the same,
elliptical galaxies can have significantly different mean stellar 
metallicities. 
Right panel shows a sequence of 
$\Delta {\rm Mg}_2/\Delta \log r=\mbox{const.}$, indicating that ellipticals
can have the same Mg$_2$ gradient while global metallicities are
different.

\subsection {Errors}

 Non-negligible scatters observed in the $\langle {\rm [Fe/H]} \rangle$ - 
$\log \sigma_0$ relation (figure 7c)
might arise from observational errors involved
in the line-strength measurements and the velocity dispersions. 
To see if one can attribute the dispersion
entirely to the observational errors, we evaluate the individual errors.

The effective radius $r_{\rm e}$ is sometimes difficult to determine,
and in the worst case, $r_{\rm e}$ differs almost 10 arcsec 
from authors to authors (Fisher, Illingworth, \& Franx 1995a).
However, with $\Delta {\rm Mg}_2 /\Delta \log r \simeq -0.051$, 
a change brought into 
mean metallicities is $\Delta {\rm [Fe/H]}\sim 0.05$, thus is negligibly
small.

It is not always the case to adopt the same spectral resolutions. 
The offset resulted from different resolutions is as large as 
$\Delta{\rm Mg}_2\sim 0.03$ and $\Delta {\rm Fe}_1 \sim 0.5$.
The resulting errors in $\langle {\rm [Fe/H]} \rangle$ are
$\Delta {\rm [Fe/H]}_{{\rm Mg}_2}\sim 0.2$ and 
$\Delta {\rm [Fe/H]}_{{\rm Fe}_1} \sim 0.3$, which could partly account 
for the dispersions, but not entirely.

Velocity dispersions in ellipticals usually decrease from the galaxy center
to the outer area. The correction factors are 1.02 for Mg$_2$, 1.23 for
Fe$_1$, 
and 1.5 for Fe$_2$ at the velocity dispersion $\sigma=300$ km s$^{-1}$, 
which are negligible for the Mg$_2$ gradient, but are significant for the
Fe$_1$ and Fe$_2$ gradients (Davies et al. 1993).

Some elliptical galaxies are known to have the surface brightness profile
that deviates significantly from de Vaucouleurs' law (Caon et al. 1993; 
Binggeli \& Jerjen 1997; Graham \& Colless 1997). Therefore, we use  
Sersic law for the surface brightness profile and study how the resulting
mean metallicity depends on the $n$ parameter. We find that 
the effect of changing $n$ is negligibly small.

Many ellipticals tend to have less steep gradients at the galaxy center.
Although it is likely that this is caused by poor seeings, one may argue
that it introduces errors in the resulting metallicity. However,
the stellar mass involved within $\log r/r_{\rm e}=-1.5$ is only 
$\sim 2\%$, and therefore the errors due to this effect should be small.

The mass-to-light ratio $M/L$ differs from galaxy to galaxy
(Faber \& Jackson 1976; Michard 1980; Schechter 1980). This would affect
the absolute magnitudes $M_B$, but not the velocity dispersions.

 None of the observational errors can fully account for the
large scatter in the $\langle {\rm [Fe/H]} \rangle$ - $\log \sigma_0$
relation.
Therefore, we conclude that the scatters in metallicity-mass relations
are at least partly real.

\subsection {Mergers?}

 To examine if this dispersion of the metallicity gradient and
the scatter of metallicity-mass relations come from galaxy mergers,
we check whether the metallicity gradients and the residuals from the
ridge line of $\langle {\rm [Fe/H]} \rangle$ - $\log \sigma_0$
relation correlate with dynamical disturbances, such as a ratio of
rotational velocity to velocity dispersion $(v/\sigma)^*$, an
isophotal contour deviation $a_4$, a dynamical peculiarity $\Sigma$,
and the mass of hot X-ray gas $M_{\rm gas}$. We fail to find any
definite evidence suggesting that the dispersion of metallicity
gradient, the scatter of metallicity-mass relations
and the dynamical disturbances come from the same origin. 

\subsection {Age Effects?}

 Major results and conclusions of this article entirely depend on our 
assumption that line-strength gradients should be read as the metallicity
gradients. The gradients of Mg$_2$, Mg$_{\rm b}$, Fe$_1$, and Fe$_2$ may also
arise from a gradient of age that decreases from the galaxy center to the
outer halo. Indeed, in many cases H$\beta$ shows a positive gradient,
thus is consistent with the age gradient scenario. To estimate possible 
influences of the age gradient on our $\langle {\rm [Fe/H]} \rangle$, we
have assumed that the Mg$_2$ and H$\beta$ gradients are caused by both 
metallicity and age, and have derived the age and metallicity gradients
simultaneously (table 3a and 3b). 
Except for the cases in which we obtain unexpectedly large
ages and/or age gradients, the resulting $\langle {\rm [Fe/H]} \rangle$
are not much different from what we obtain with no age gradients,
and the scatter of metallicity-mass relation is nearly the same
even if age gradients are considered.
However, H$\beta$ is not a good age indicator, and until it becomes possible 
to analyze the gradients of new H$\gamma$ defined by 
Vazdekis \& Arimoto (1999), 
one cannot exclude the possibility of age gradient conclusively.

 Nevertheless, we wish to claim that the line-strength gradients are due to
the metallicity gradient. Quite recently, Tamura et al. (1999) have measured
gradients of HST V$_{606}$-I$_{814}$ colors of 7 elliptical-like galaxies
in the Hubble Deep Field (HDF); 2-456.0 (z=0.089), 2-121.0 (z=0.475), 
3-790.0 (z=0.562), 3-321.0 (z=0.677), 4-744.0 (z=0.764), 4-493.0 (z=0.847),
and 4-752.0 (z=1.013). These galaxies are red and their luminosity 
profiles follow de Vaucouleurs' law, thus it is very likely that they are
genuine ellipticals. The V$_{606}$-I$_{814}$ gradients of these galaxies
do not show significant difference, and this means that the color gradients of
ellipticals evolve passively since $z \simeq 1$. Tamura et al. (1999) build
two evolutionary sequences of ellipticals by assuming that (a) the color 
gradients are entirely due to the metallicity gradient (a metallicity
sequence)
and (b) the color gradients arise from the age gradient (an age sequence).
The color gradients of the metallicity sequence evolve passively and do not
change significantly from the present to $z \simeq 1$. On the other hand, the
color gradients of the age sequence become significantly steeper beyond
$z\simeq 0.5$ and become almost vertical at $z\simeq 1$. Obviously, 7
ellipticals
in the HDF do not agree with an evolution of the age sequence at all, 
but show quite
excellent agreement with theoretical evolutionary loci of the metallicity 
sequence. Therefore, the color gradients of ellipticals certainly reflect the
gradient of stellar metallicity. Then, why not the line-strength gradients? 

\section {Conclusions}

We have re-studied line-strength gradients of 80 elliptical galaxies. 
Our sample galaxies show a very similar central (Mg$_2$)$_0$
distribution to that of 572 ellipticals studied by Davies et al. (1987).
We assume that the gradients of metallic lines, such as
Mg$_2$, Mg$_{\rm b}$, Fe$_1$, and Fe$_2$, 
all originate from the gradient of mean stellar
metallicity in elliptical galaxies,
and have applied Worthey's (1994) index-metallicity relations
to get $\Delta {\rm [Fe/H]} / \Delta \log r=-0.30\pm0.12$ (Mg$_2$), 
$-0.34\pm0.16$ (Mg$_{\rm b}$),
$-0.28\pm0.16$ (Fe$_1$), and $-0.25\pm0.13$ (Fe$_2$)
and find more than 80\% of elliptical
galaxies show the metallicity gradients in the range of 
$-0.6 \le \Delta {\rm [Fe/H]}/ \Delta \log r \le -0.1$.
A typical gradient $\Delta {\rm [Fe/H]}/ \Delta \log r \simeq -0.3$
is flatter than 
the gradients predicted by monolithic collapse simulations
(Larson 1974a; Carlberg 1984).
The metallicity gradients do not correlate with any physical properties of
galaxies, including central and mean metallicities, 
central velocity dispersions $\sigma_0$ , absolute B-magnitudes $M_B$, 
absolute effective radii $R_{\rm e}$, and dynamical masses of galaxies.
Unless there are significant unknown errors in the measurements of 
line-strength gradients, elliptical galaxies have
different metallicity gradients, even if they have nearly identical
properties such as masses and luminosities. This rather surprising behavior
of the line-strength gradients has never been taken 
into account in modeling the formation of elliptical galaxies. 

By using the metallicity gradients, we have calculated mean stellar 
metallicities
for individual ellipticals. The average metallicities of our sample are
$\langle {\rm [Fe/H]}\rangle =-0.18\pm0.19$ (Mg$_2$), 
$-0.15\pm0.24$ (Mg$_{\rm b}$),
$-0.37\pm0.17$ (Fe$_1$), and $-0.36\pm0.18$ (Fe$_2$),
and a typical metallicity is a half solar,
$\langle {\rm [Fe/H]} \rangle \simeq -0.3$.
Contrary to what
Gonzalez \& Gorgas (1996) claimed, 
the mean metallicities of ellipticals are not universal, 
but range from $\langle {\rm [Fe/H]} \rangle \simeq -0.8$ to $+0.3$.
Since the mean metallicity of 
19 elliptical galaxies with $\sigma_0 \ge 250$ km s$^{-1}$
is $\langle {\rm [Fe/H]} \rangle \simeq -0.04 \pm 0.11$,
a claim made by Arimoto et al. (1997) remains unexplained; i.e.,
the iron abundance of the interstellar mediums of luminous elliptical
galaxies, as derived from ASCA X-ray observations of the iron L complex,
is at variance with the abundance expected from the stellar populations. 
$\langle {\rm [Fe/H]} \rangle$ correlates
well with $\sigma_0$ and dynamical mass, 
though relations for M$_B$ and R$_{\rm e}$ include significant scatters.
We find the fundamental planes defined by surface brightnesses SB$_{\rm e}$, 
$\langle {\rm [Fe/H]} \rangle$, and $R_{\rm e}$ (or $M_B$),
the scatters of which are much smaller than those of the 
$\langle {\rm [Fe/H]} \rangle$ - $R_{\rm e}$ (or $M_B$) relation.
The $\langle [{\rm Fe/H]} \rangle$ - $\log \sigma_0$ 
relation is nearly in parallel to the [Fe/H]$_0$ - $\log\sigma_0$ 
relation but systematically lower by 0.3 dex; thus 
the mean metallicities are about a half of the central values. 
The metallicity-mass relation, or equivalently, the color-magnitude relation
of ellipticals holds not only for the central part but also for the whole
part of galaxies.

 Assuming that Mg$_2$ and Fe$_1$ give [Mg/H] and [Fe/H], respectively, 
we find $\langle {\rm [Mg/Fe]} \rangle \simeq +0.2$ in most of 
elliptical galaxies. 
$\langle {\rm [Mg/Fe]} \rangle$ shows no correlation with galaxy mass tracers
such as $\sigma_0$, 
in contrast to what was claimed for the central [Mg/Fe]. 
This can be most naturally explained if the star formation had stopped 
in elliptical galaxies before the bulk of SNe Ia began to explode.

  Elliptical galaxies can have significantly different
metallicity gradients and $\langle {\rm [Fe/H]} \rangle$ even if
they have the same galaxy mass. This may result from different
history of galaxy mergers, as suggested by White's (1980)
simulation. However, we fail to find any evidence suggesting
the same origin for the dispersion of metallicity gradient,
the scatter around the metallicity-mass relation, and the dynamical
disturbances; none of available data of $\Sigma$,
$a_4$, and $(v/\sigma)^*$ show a correlation with
the gradients and the scatters.
The scatters of both gradient and metallicity-mass relation
might come from dust obscuration, but contrary to broad band colors,
line strengths are far less sensitive to the dust.
Another possibility is that the scatters are due to
different ages and age gradients. However, the scatters are affected
little by the age and age gradient which we derive from H$_{\beta}$,
although H$_{\beta}$ may not be a good age indicator.
If it turns out that these galaxies
are old and have no peculiar age gradients, the scatters discussed
in this paper should have their origin at the formation epoch of galaxies.

\acknowledgments

We would like to express our heartfelt thanks to our anonymous referee
whose critical reading of our draft improved significantly the paper. 
We are grateful to C.M.Carollo, J.Gorgas, and R.P.Saglia for providing us
the machine readable data of their line-strength gradients. We thank to 
T.Yamada, K.Ohta and N.Tamura for fruitful discussions. 
C.K. thanks to the Japan Society for Promotion 
of Science for a finantial support.
This work was financially 
supported in part by a Grant-in-Aid for the Scientific Research (No.0940311)
by the Japanese Ministry of Education, Culture, Sports and Science.

\appendix
\section{Mg$_2$ and Mg$_{\rm b}$ gradients of early-type galaxies}

To give a comprehensive view of the line-strength gradients, 
we show in figures 12 and 13 the Mg$_2$ and Mg$_{\rm b}$ gradients 
of 133 early-type galaxies
including compact ellipticals, S0 galaxies, and bulge of spirals.
Filled circles and crosses 
are the observational data taken from F77, EG, CH, PEL, GEA,
BT, D91a, D91b, BS, D92, CDB, DSP, GON, HP, SAG, CDa, CDb, SPD, CGA,
and FFI.
We plot the data taken along several position angles,
after correcting for the eccentricity.
Solid lines give regression lines which are calculated by
using only filled circles.

{}

\newpage

\begin{figure}[ht]
\figurenum{1}
\centerline{\psfig{figure=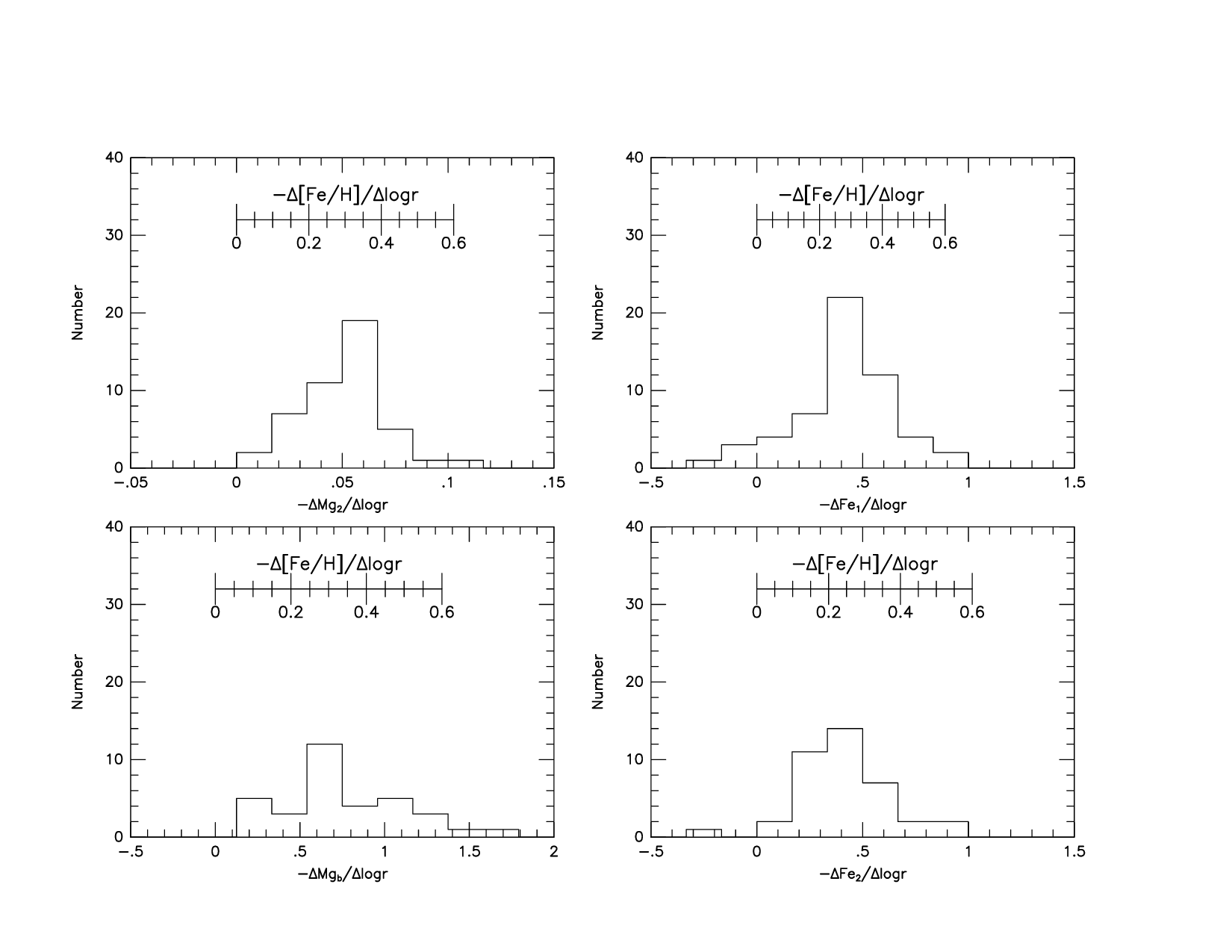,width=14cm}}
\caption[kobayashi_fig1.ps]{
The frequency distributions of the line-strength gradients. 
From the upper-left to the lower-left in clockwise, the panels 
show Mg$_2$, Fe$_1$, Fe$_2$, and Mg$_{\rm b}$, respectively. 
The metallicity scales are taken from the
index-metallicity relations of Worthey's (1994) population 
synthesis models, provided that galaxies are 17 Gyr old.
\label{fig1}}
\end{figure}

\begin{figure}[ht]
\figurenum{2}
\centerline{\psfig{figure=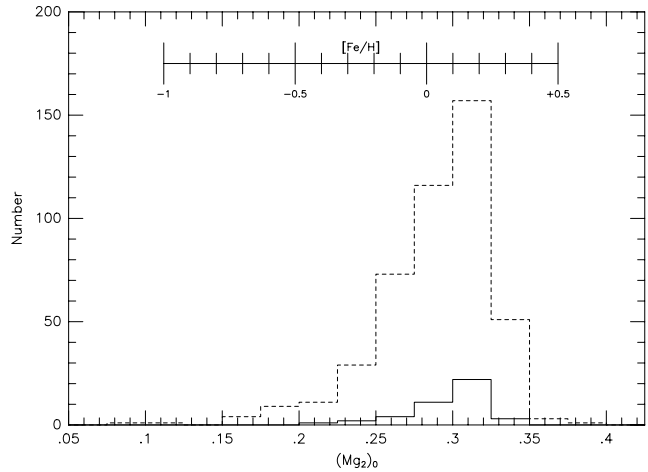,width=10cm}}
\caption[kobayashi_fig2.ps]{
The frequency distribution of Mg$_2$ at a galaxy center  
for 572 ellipticals (dashed line) taken from Davies et al. (1987)
and 46 ellipticals of our sample (solid line).
The metallicity scale is the same as in figure 1.
\label{fig2}}
\end{figure}

\begin{figure}[ht]
\figurenum{3}
\centerline{\psfig{figure=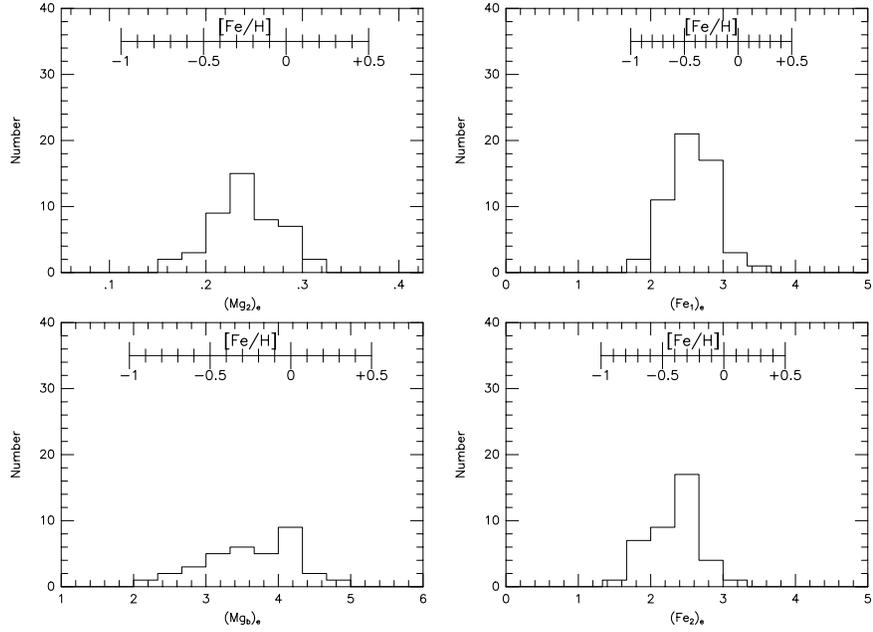,width=14cm}}
\caption[kobayashi_fig3.ps]{
The same as figure 1, but for the intercepts at the effective radius.
\label{fig3}}
\end{figure}

\begin{figure}[ht]
\figurenum{4}
\centerline{\psfig{figure=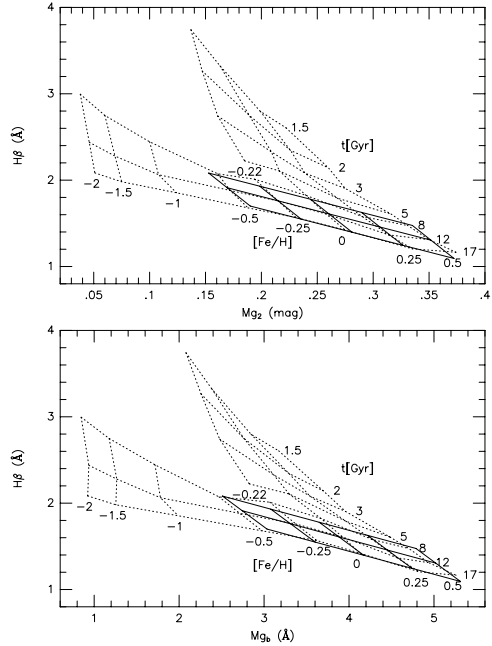,width=8cm}}
\caption[kobayashi_fig4.ps]{
H$\beta$ versus Mg$_2$ (upper panel) and Mg$_{\rm b}$ (lower panel) relations
as a function of age and metallicity.
Dotted lines give simple stellar population models of Worthey (1994),
and solid lines are the approximated relations adopted in this paper.
\label{fig4}}
\end{figure}

\begin{figure}[ht]
\figurenum{5}
\centerline{\psfig{figure=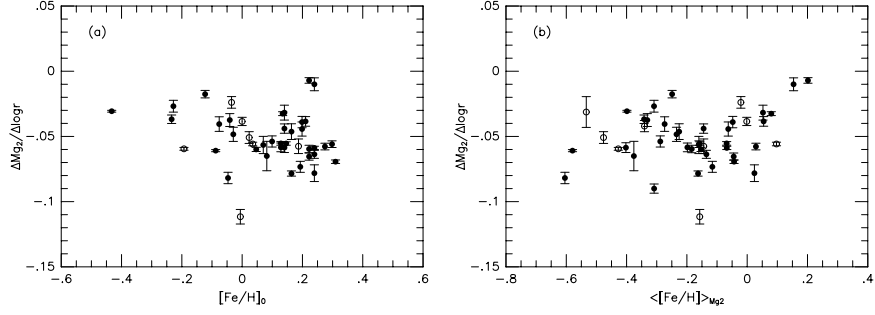,width=14cm}}
\caption[kobayashi_fig5.ps]{
The Mg$_2$ gradient versus the metallicity.
Open and filled circles show cD and elliptical galaxies, respectively. 
(a) and (b) are for the central metallicity, 
and the mean stellar metallicity calculated 
with equation (22), respectively.
\label{fig5}}
\end{figure}

\begin{figure}[ht]
\vspace*{-4cm}
\figurenum{6}
\centerline{\psfig{figure=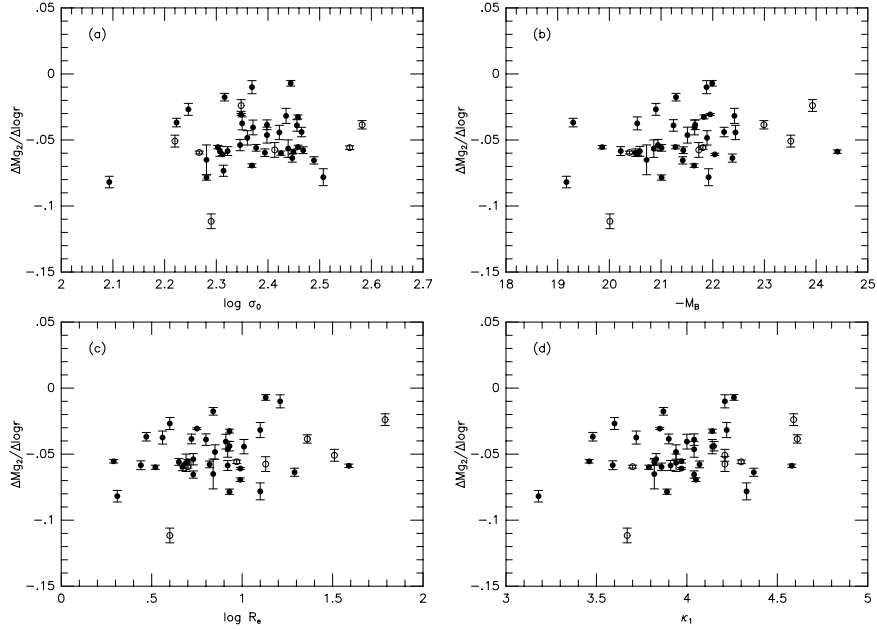,width=14cm}}
\caption[kobayashi_fig6.ps]{
The Mg$_2$ gradient versus the mass tracers; 
(a) the central velocity dispersion $\sigma_0$,
(b) the absolute B-magnitude $M_B$, 
(c) the absolute effective radius $R_{\rm e}$, and
(d) the dynamical mass $\kappa_1=(\log\sigma_0^2+\log R_{\rm e})/\sqrt{2}$.
Symbols are the same as in figure 5.
\label{fig6}}
\end{figure}

\begin{figure}[ht]
\figurenum{7}
\centerline{\psfig{figure=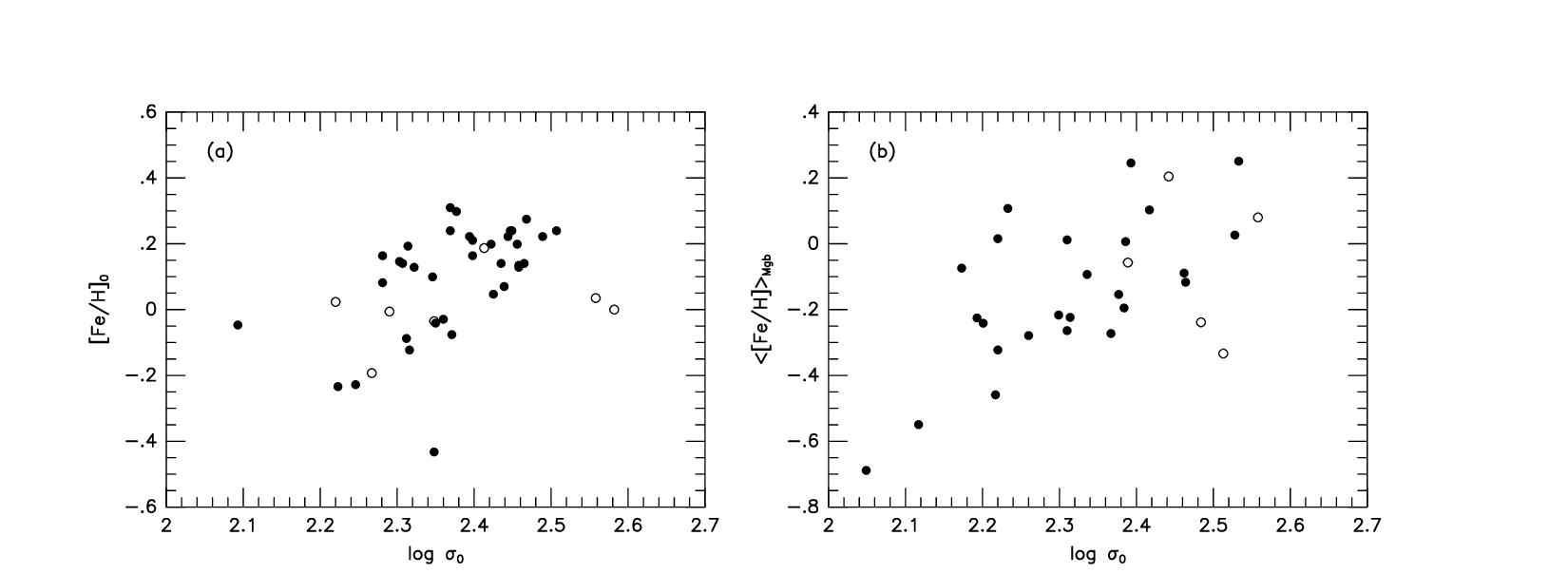,width=14cm}}
\centerline{\psfig{figure=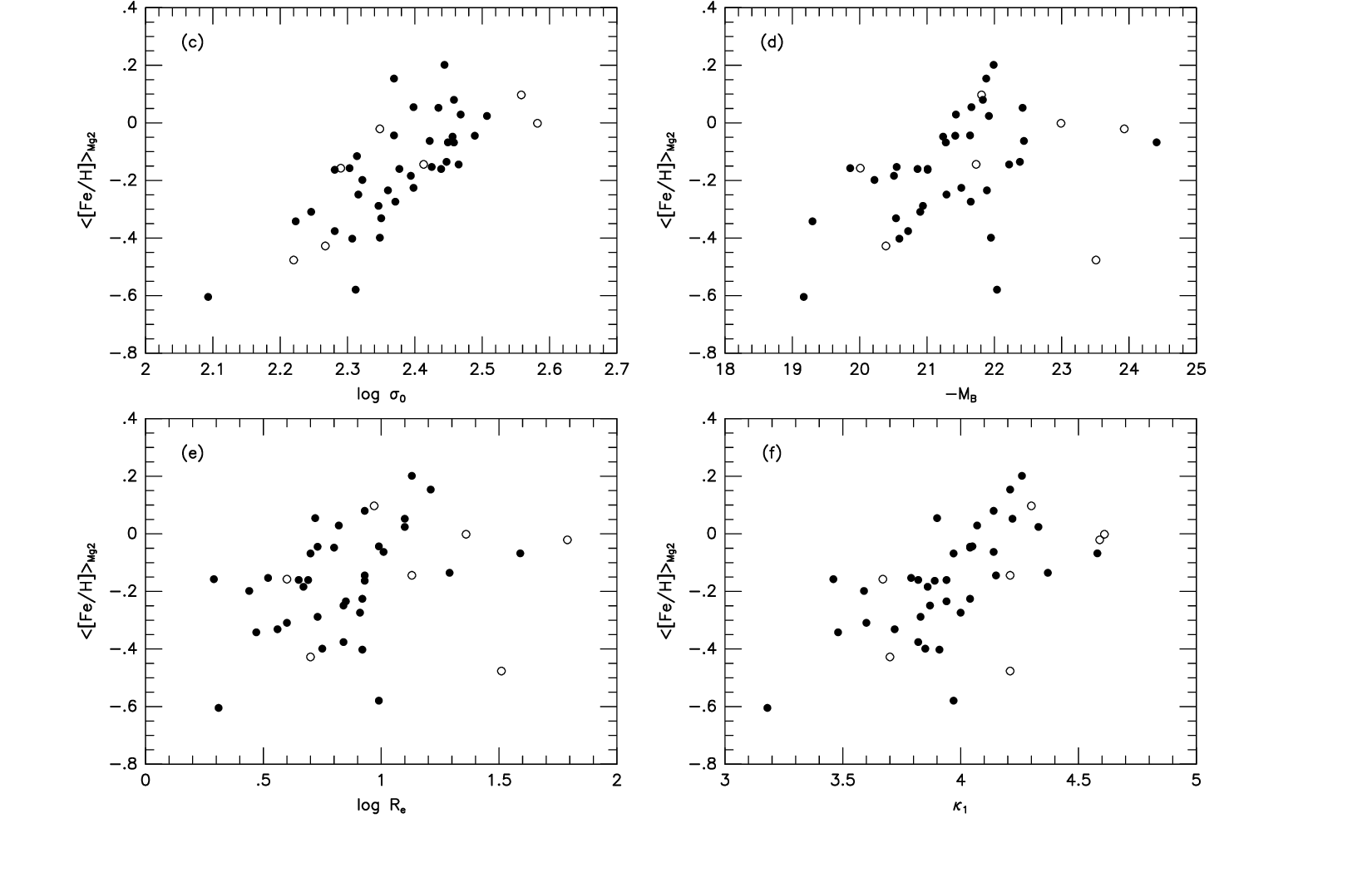,width=14cm}}
\caption[kobayashi_fig7.ps]{
The metallicity versus the mass tracers; 
(a)-(c) $\sigma_0$, (d) $M_B$, (e) $R_{\rm e}$, and (f) $\kappa_1$.
(a) is for the central metallicity derived from the central Mg$_2$ index,
(b) is for the mean stellar metallicity derived from the Mg$_{\rm b}$ gradient,
and (c)-(f) are for the mean stellar metallicity derived from the Mg$_2$ gradient.
Symbols are the same as in figure 5.
\label{fig7}}
\end{figure}

\begin{figure}[ht]
\figurenum{8}
\centerline{\psfig{figure=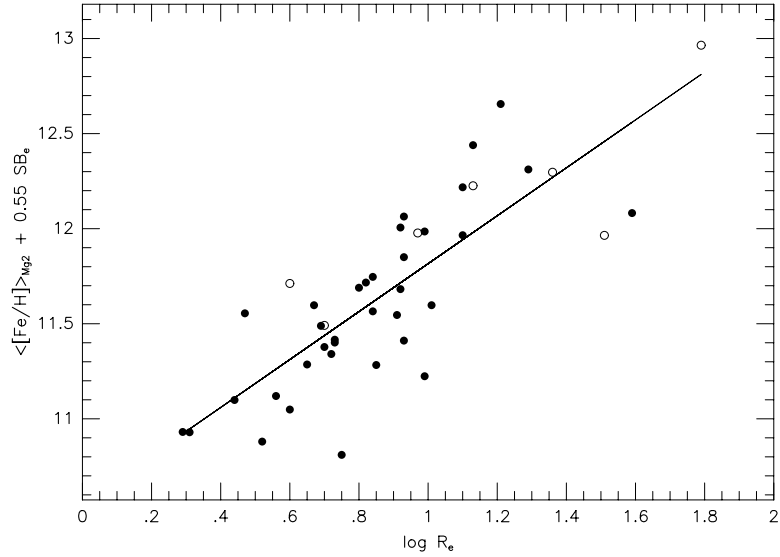,width=12cm}}
\caption[kobayashi_fig8.ps]{
The fundamental plane for the absolute radius, 
the surface brightness, and the mean metallicity.
The symbols are the same as in figure 5.  
\label{fig8}}
\end{figure}

\begin{figure}[ht]
\vspace*{-3cm}
\figurenum{9}
\centerline{\psfig{figure=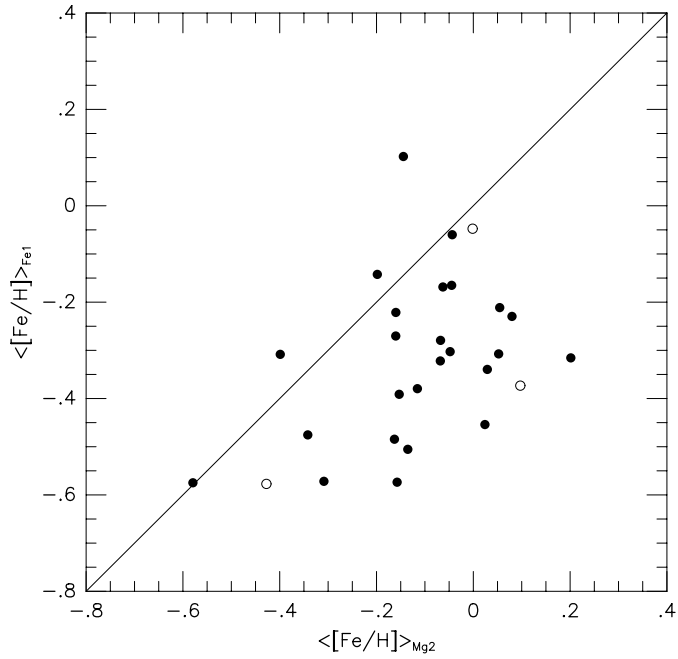,width=14cm}}
\caption[kobayashi_fig9.ps]{
Comparison of the mean metallicities derived from Mg$_2$ and Fe$_1$. 
The symbols are the same as in figure 5.  
\label{fig9}}
\end{figure}

\begin{figure}[ht]
\figurenum{10}
\centerline{\psfig{figure=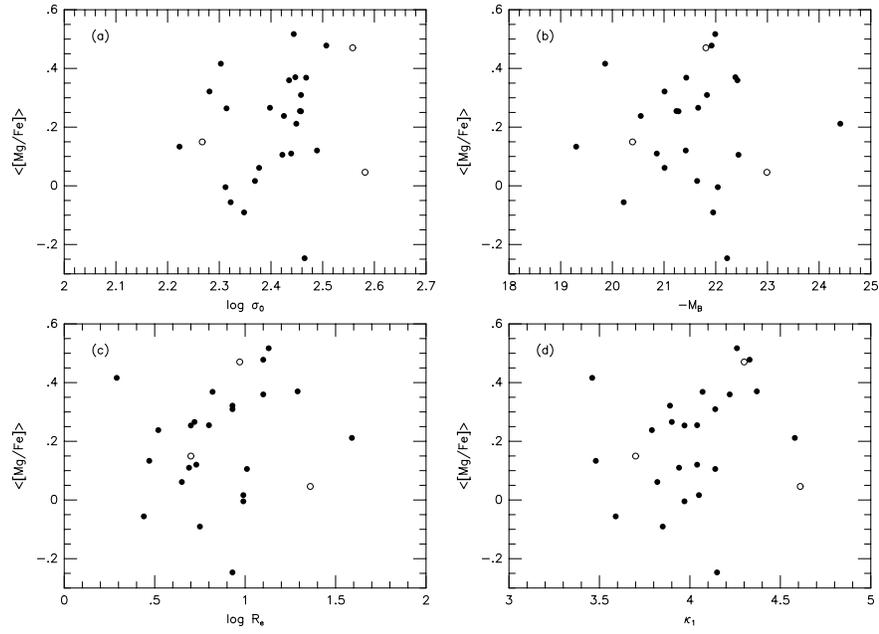,width=14cm}}
\caption[kobayashi_fig10.ps]{
The mean [Mg/Fe] ratio versus the mass tracers; 
(a) $\sigma_0$, (b) $M_B$, (c) $R_{\rm e}$, and (d) $\kappa_1$.
Symbols are the same as in figure 5.
\label{fig10}}
\end{figure}

\begin{figure}[ht]
\figurenum{11}
\centerline{\psfig{figure=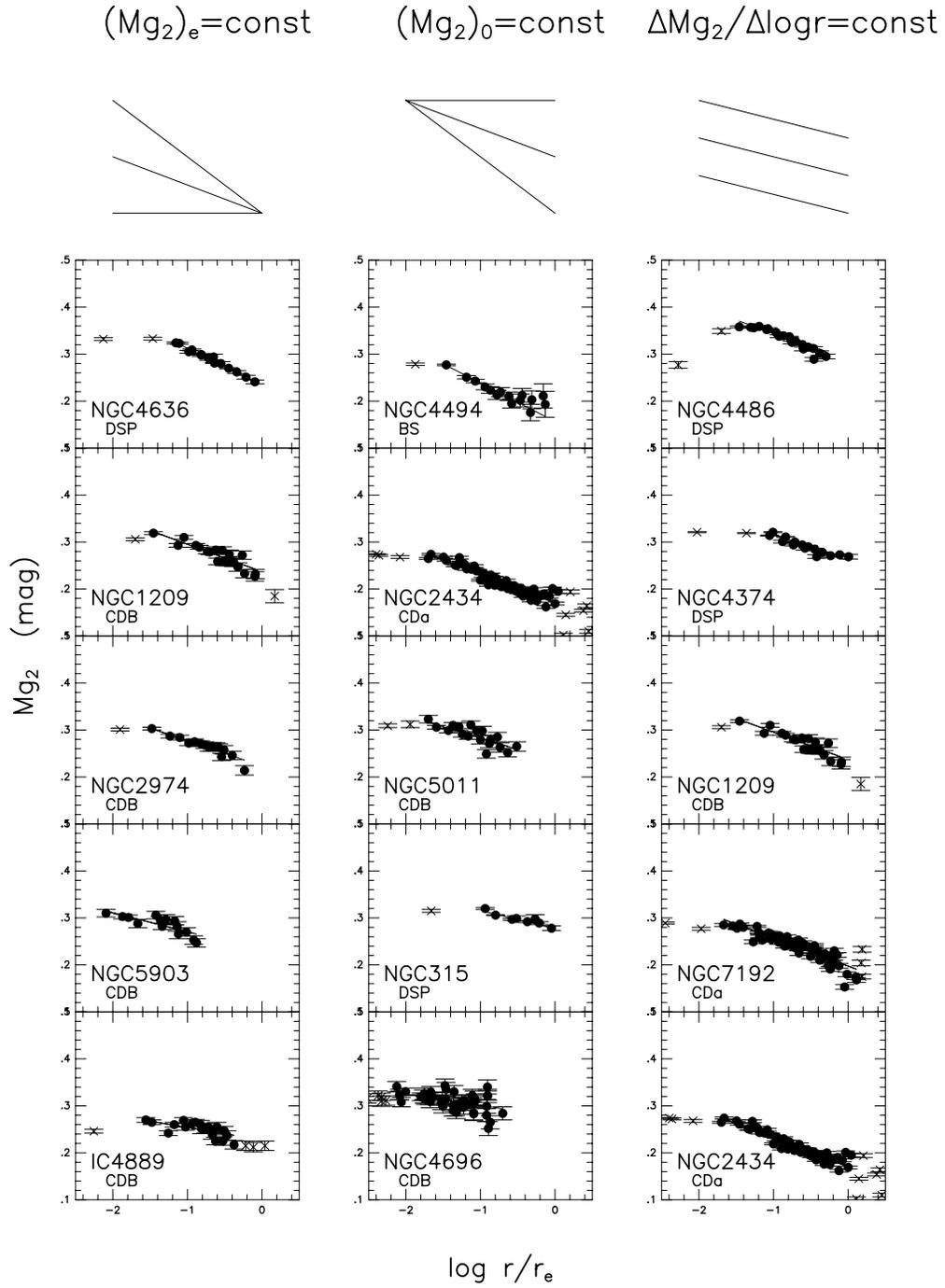,width=16cm}}
\caption[kobayashi_fig11.ps]{
The Mg$_2$ gradients of elliptical galaxies. 
The left, middle, and right panels give the
sequences of $\left({\rm Mg}\right)_{\rm e}=\mbox{const.}$,
$\log \sigma_0=\mbox{const.}$, and
$\Delta{\rm Mg}_2/\Delta\log r=\mbox{const.}$, respectively.
\label{fig11}}
\end{figure}

\begin{figure}[ht]
\figurenum{12}
\centerline{\psfig{figure=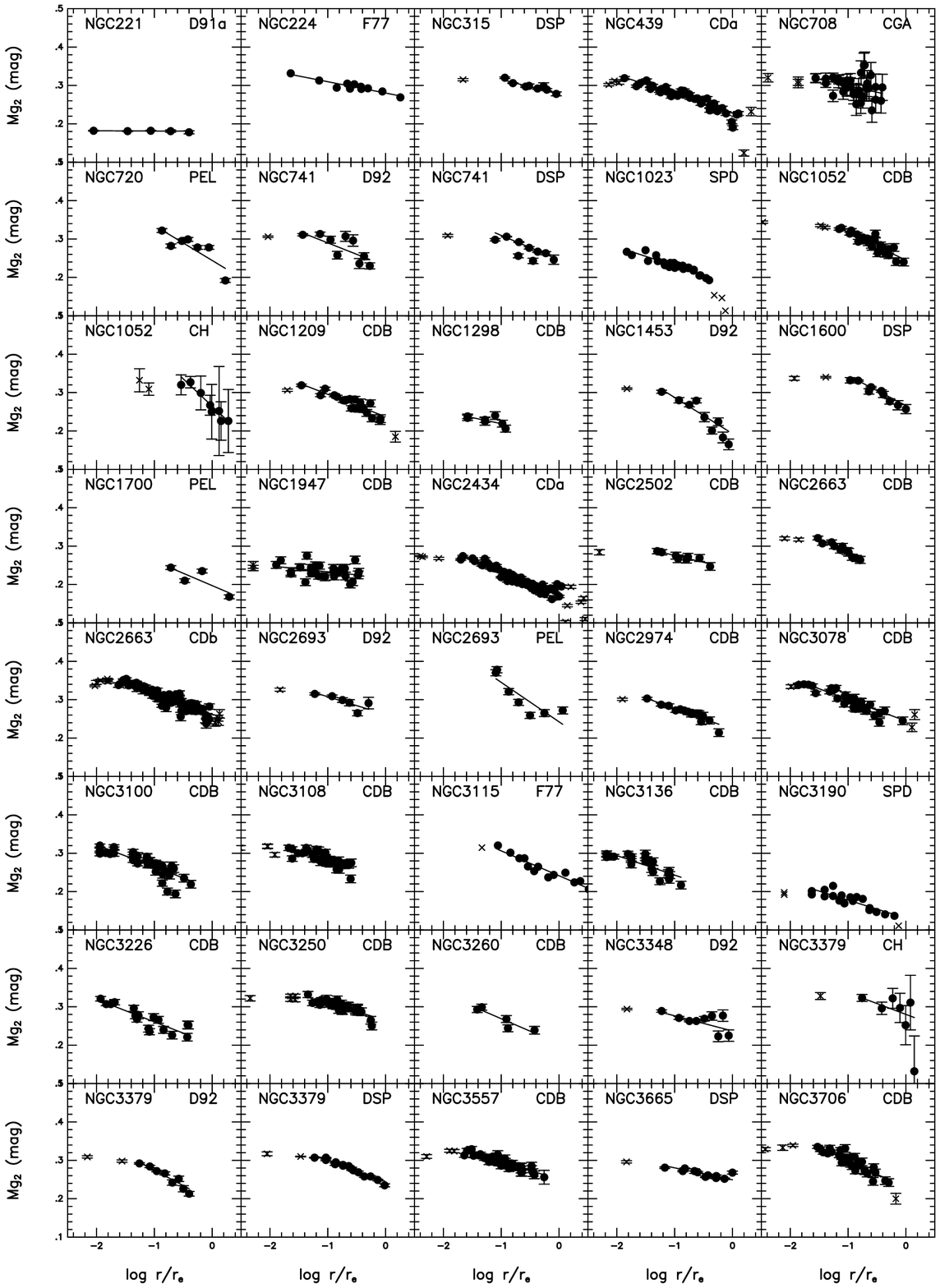,width=18cm}}
\caption[kobayashi_fig12.ps]{
The Mg$_2$ gradients of early-type galaxies. 
\label{fig12}}
\end{figure}
\begin{figure}[ht]
\centerline{\psfig{figure=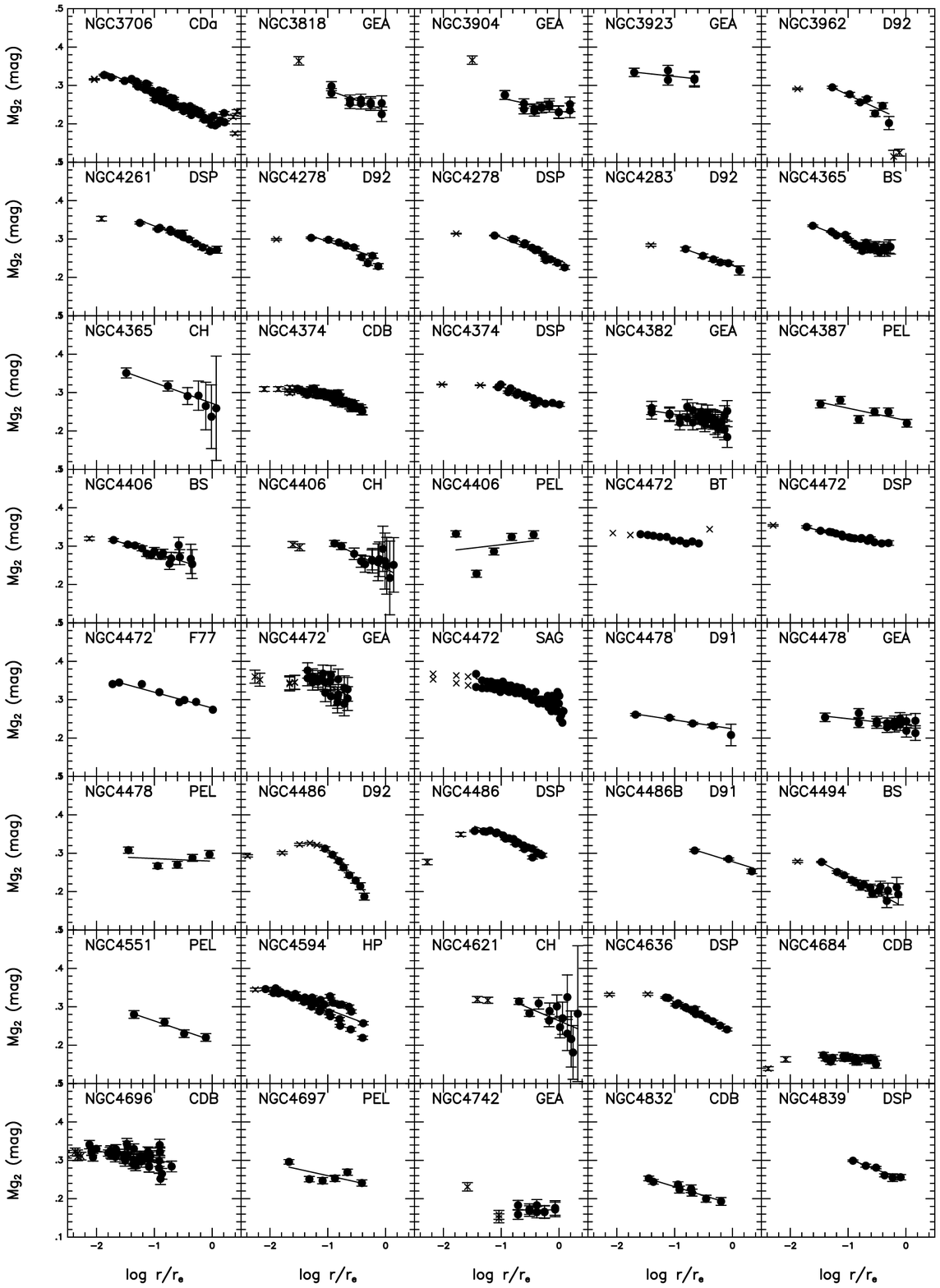,width=18cm}}
\end{figure}
\begin{figure}[ht]
\centerline{\psfig{figure=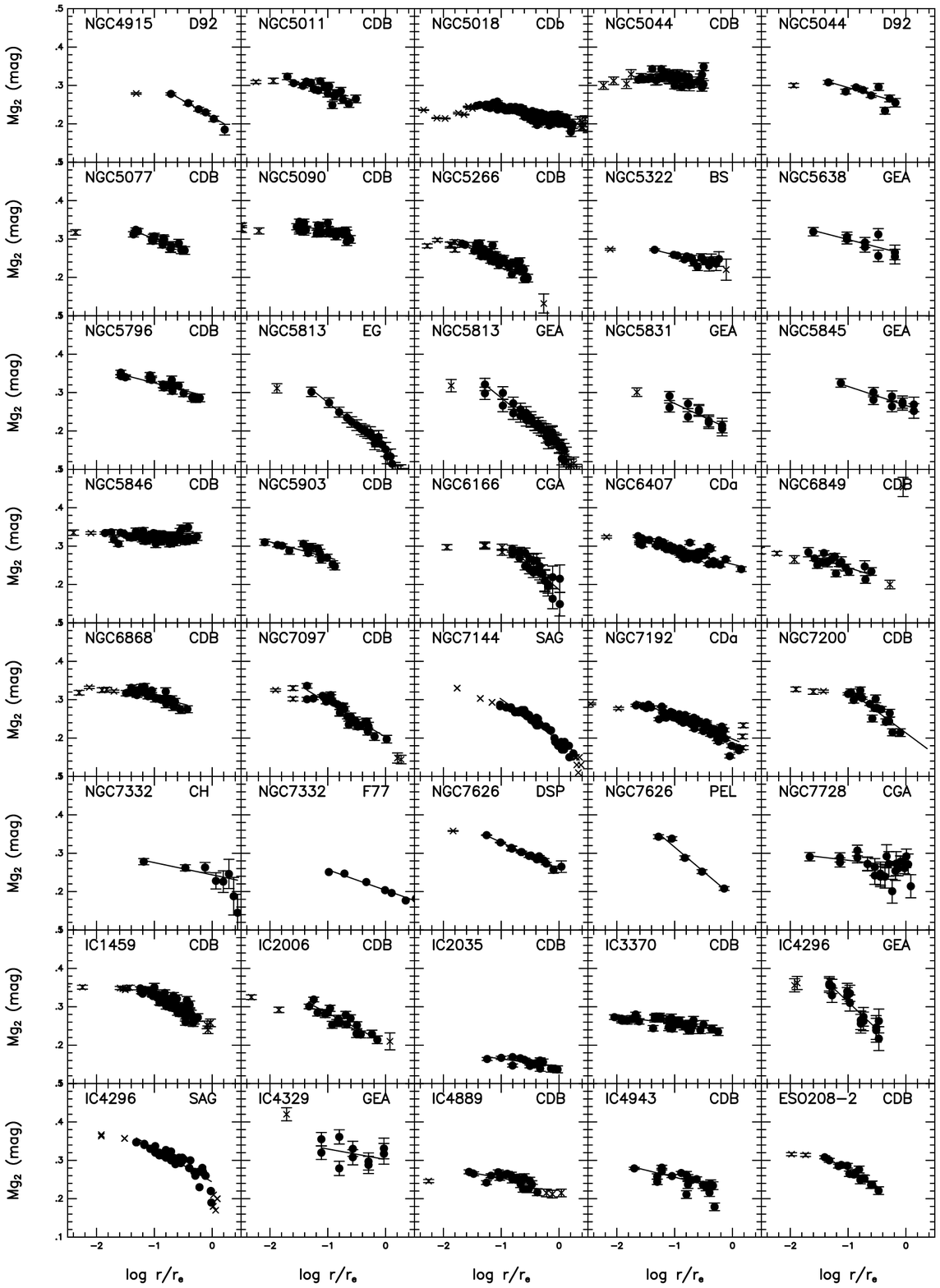,width=18cm}}
\end{figure}
\begin{figure}[ht]
\centerline{\psfig{figure=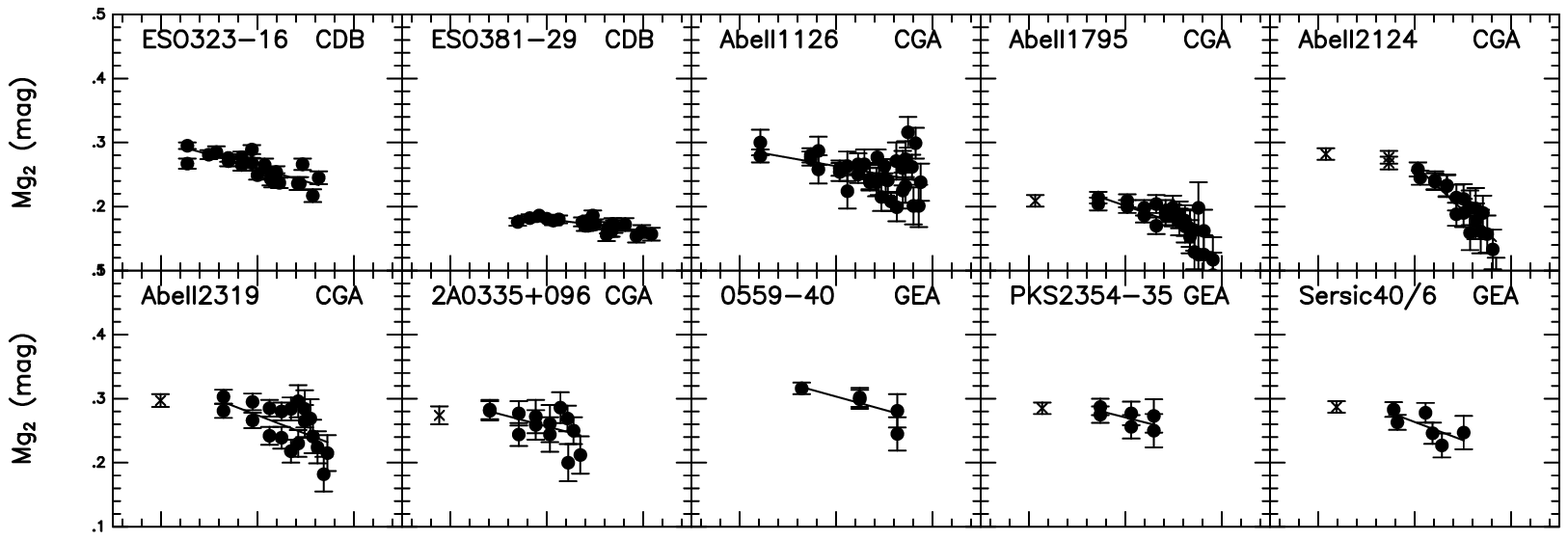,width=18cm}}
\end{figure}

\begin{figure}[ht]
\figurenum{13}
\centerline{\psfig{figure=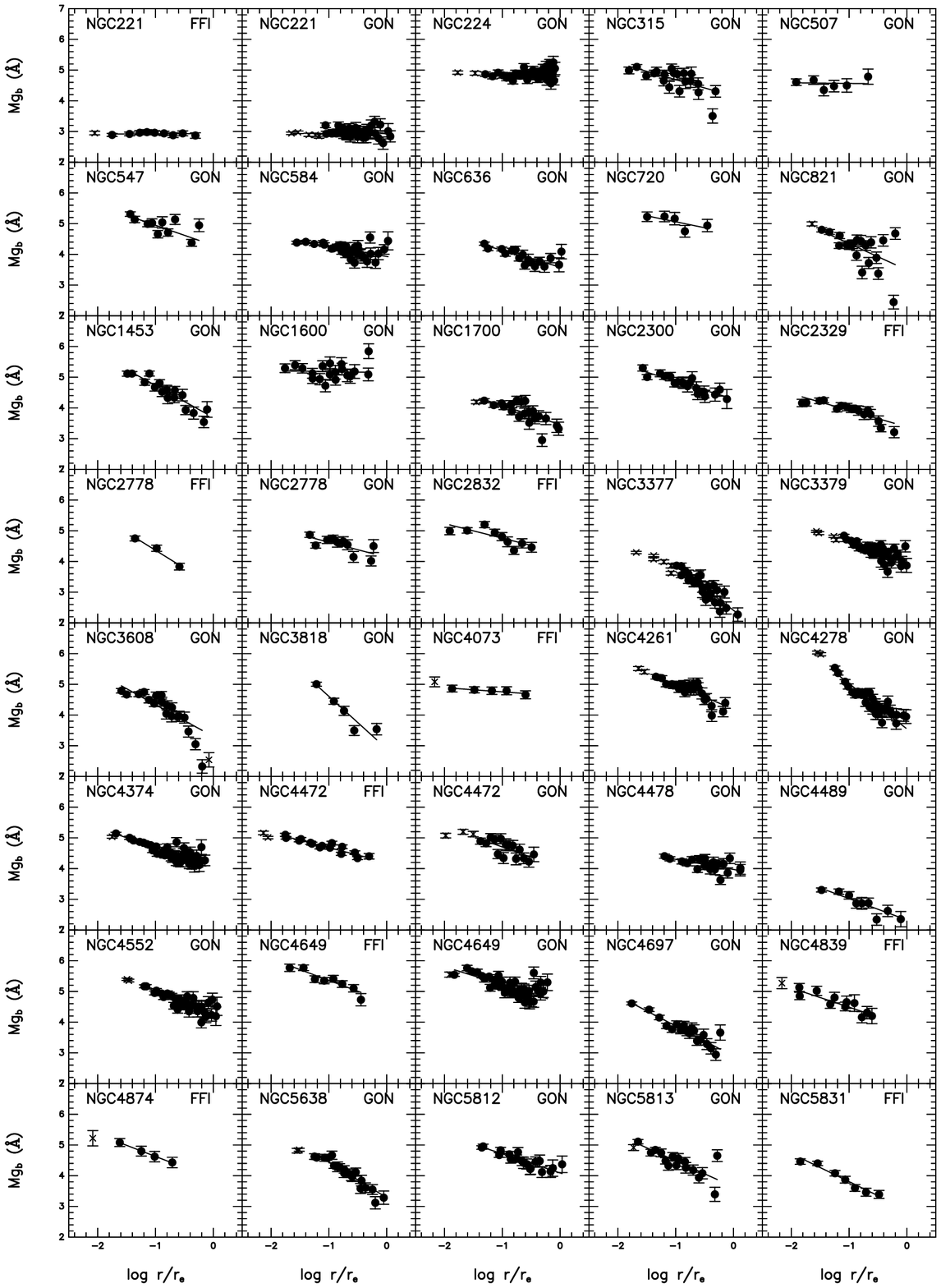,width=18cm}}
\caption[kobayashi_fig13.ps]{
The Mg$_{\rm b}$ gradients of early-type galaxies. 
\label{fig13}}
\end{figure}
\begin{figure}[ht]
\centerline{\psfig{figure=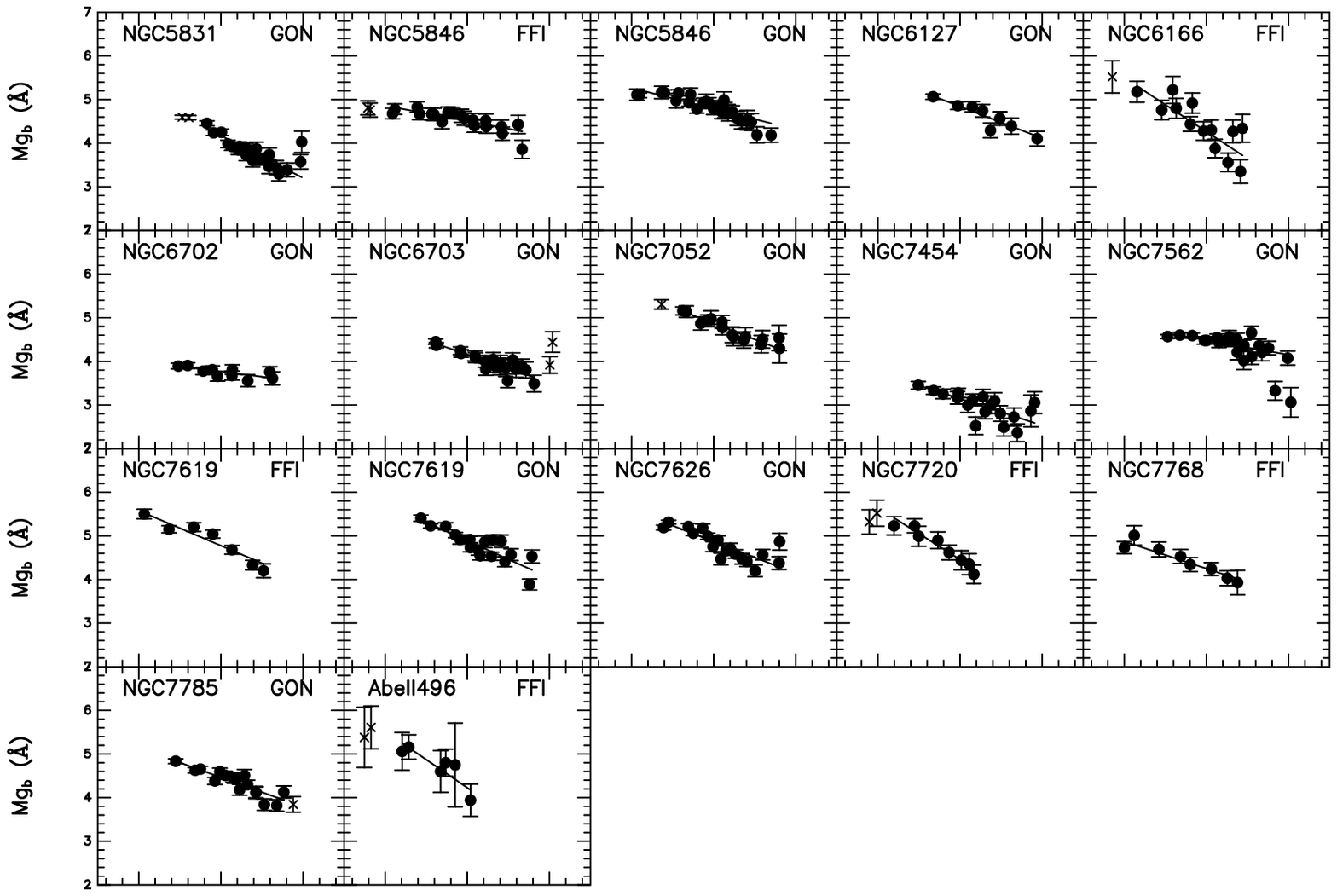,width=18cm}}
\end{figure}

\newpage



\begin{deluxetable}{llrrl}
\tablenum{5}
\footnotesize
\tablewidth{0pt}
\tablecaption{Global Scaling Relations \label{tab5}}
\tablehead{
\colhead{$x$} & \colhead{$y$} & \colhead{N} &
\colhead{$r$} & \colhead{Regression Line}
}
\startdata
${\rm [Fe/H]}_0$                  & $\Delta{\rm Mg}_2/\Delta\log r$ 
  & 43  & $-0.21$ &  \nl
$\langle {\rm [Fe/H]} \rangle$    & $\Delta{\rm Mg}_2/\Delta\log r$
  & 46  & $0.22$ &  \nl
$\log\sigma_0$                    & $\Delta{\rm Mg}_2/\Delta\log r$
  & 43  & $0.17$ &  \nl
$\log R_{\rm e}$                  & $\Delta{\rm Mg}_2/\Delta\log r$
  & 42  & $0.26$ &  \nl
$\kappa_1$                        & $\Delta{\rm Mg}_2/\Delta\log r$
  & 42  & $0.27$ &  \nl
$M_B$                             & $\Delta{\rm Mg}_2/\Delta\log r$
  & 42  & $0.33$ &  \nl
$\log\sigma_0$                    & ${\rm [Fe/H]}_0$  
  & 43  & $ 0.45$      & $y=0.76  \,x  -1.73$     \nl
$\log\sigma_0$                    & $\langle {\rm [Fe/H]} \rangle$ 
  & 43  & $ 0.74$ & $y=1.46  \,x  -3.62$     \nl
$\log R_{\rm e}$                  & ${\rm [Fe/H]}_0$ 
  & 42  & $ 0.22$ &  \nl
$\log R_{\rm e}$                  & $\langle {\rm [Fe/H]} \rangle$
  & 42  & $ 0.33$ &  \nl
$\kappa_1$                        & ${\rm [Fe/H]}_0$
  & 42  & $ 0.37$      & $y=0.20  \,x  -0.71$     \nl
$\kappa_1$                        & $\langle {\rm [Fe/H]} \rangle$
  & 42  & $ 0.59$ & $y=0.37  \,x  -1.62$     \nl
$M_B$                             & ${\rm [Fe/H]}_0$
  & 42  & $ 0.19$ &  \nl
$M_B$                             & $\langle {\rm [Fe/H]} \rangle$
  & 42  & $ 0.37$ &  \nl
$\log\sigma_0$                    & $\langle {\rm [Mg/Fe]} \rangle$
  & 27  & $ 0.22$ &  \nl
$\log R_{\rm e}$                  & $\langle {\rm [Mg/Fe]} \rangle$
  & 26  & $ 0.15$ &  \nl
$\kappa_1$                        & $\langle {\rm [Mg/Fe]} \rangle$
  & 26  & $ 0.00$ &  \nl
$M_B$                             & $\langle {\rm [Mg/Fe]} \rangle$
  & 26  & $ 0.21$ &  
\enddata
\end{deluxetable}

\begin{deluxetable}{lllrl}
\tablenum{6}
\footnotesize
\tablewidth{0pt}
\tablecaption{Fundamental Planes \label{tab6}}
\tablehead{
\colhead{$x$} & \colhead{$y$} & \colhead{$z$} &
\colhead{N} & \colhead{Regression Line}
}
\startdata
$\log\sigma_0$
 & $\log R_{\rm e}$
 & ${\rm SB}_{\rm e}$
 & 42 & $x-0.64\,y+0.27\,z =7.70$  \nl
$\log\sigma_0$
 & $M_B$
 & ${\rm SB}_{\rm e}$
 & 42 & $x-0.09\,y+0.12\,z =3.17$  \nl
$\left({\rm Mg}_2\right)_0$
 & $\log R_{\rm e}$
 & ${\rm SB}_{\rm e}$
 & 42 & $x-0.26\,y+0.13\,z =2.80$  \nl
$\left({\rm Mg}_2\right)_0$
 & $M_B$
 & ${\rm SB}_{\rm e}$
 & 42 & $x-0.04\,y+0.09\,z =1.41$  \nl
$\langle {\rm [Fe/H]} \rangle$
 & $\log R_{\rm e}$
 & ${\rm SB}_{\rm e}$
 & 42 & $x-1.26\,y+0.55\,z =10.56$  \nl
$\langle {\rm [Fe/H]} \rangle$
 & $M_B$
 & ${\rm SB}_{\rm e}$
 & 42 & $x-0.16\,y+0.28\,z =2.41$  \nl
\enddata
\end{deluxetable}

\end{document}